\documentclass[fleqn,usenatbib]{mnras}

\usepackage[T1]{fontenc}

\DeclareRobustCommand{\VAN}[3]{#2}
\let\VANthebibliography\thebibliography
\def\thebibliography{\DeclareRobustCommand{\VAN}[3]{##3}\VANthebibliography}

\usepackage{graphicx}	% Including figure files
\usepackage{amsmath,bm}	% Advanced maths commands
\usepackage{amssymb}	% Extra maths symbols
\usepackage{xspace}
\usepackage{multirow}
\usepackage{savesym} % Fix for siunitx causing a clash with \tablenum
\savesymbol{tablenum}
\usepackage{siunitx}
\restoresymbol{SIX}{tablenum}
\usepackage[normalem]{ulem}

\usepackage{newtxtext,newtxmath}
\newcommand{\vect}[1]{\ensuremath{\bm{#1}}}
\newcommand{\matr}[1]{\ensuremath{\bm{\mathrm{#1}}}}
\newcommand{\Planck}{\textit{Planck}\xspace}
\newcommand{\Swift}{\textit{Swift}\xspace}
\newcommand{\colheadd}[1]{{\bf #1}}
\newenvironment{closetabcols}[1][0.5mm]{\setlength{\tabcolsep}{#1}}{}

\title[ACT: targeted transients]{The Atacama Cosmology Telescope: Flux Upper Limits from a Targeted Search for Extragalactic Transients}

\author[C.~Herv\'ias-Caimapo~et~al.]{
Carlos~Herv\'ias-Caimapo,$^{1,2}$\thanks{E-mail:carlos.hervias@uc.cl}
Sigurd~Naess,$^{3}$
Adam~D.~Hincks,$^{4,5}$
Erminia~Calabrese,$^{6}$
Mark~J.~Devlin,$^{7}$
\newauthor
Jo~Dunkley,$^{8,9}$
Rolando~D{\"u}nner,$^{2}$
Patricio~A.~Gallardo,$^{10}$
Matt~Hilton,$^{11,12}$
Anna~Y.~Q.~Ho,$^{13}$
\newauthor
Kevin~M.~Huffenberger,$^{1}$
Xiaoyi~Ma,$^{14}$
Mathew~S.~Madhavacheril,$^{7}$
Michael~D.~Niemack,$^{13,15}$
\newauthor
John~Orlowski-Scherer,$^{7}$
Lyman~A.~Page,$^{8}$
Bruce~Partridge,$^{16}$
Roberto~Puddu,$^{2}$
Maria~Salatino,$^{17,18}$
\newauthor
Crist\'obal~Sif\'on,$^{19}$
Suzanne~T.~Staggs,$^{20}$
Cristian~Vargas,$^{2}$
Eve~M.~Vavagiakis,$^{15}$
Edward~J.~Wollack$^{21}$
\newauthor
\newline
\textit{\normalsize Affiliations are listed at the end of the paper.}%
}%

\date{Accepted XXX. Received YYY; in original form ZZZ}

\pubyear{2023}

\begin{document}
\label{firstpage}
\pagerange{\pageref{firstpage}--\pageref{lastpage}}
\maketitle

\begin{abstract}
We have performed targeted searches of known extragalactic transient events at millimetre wavelengths using nine seasons (2013--2021) of 98, 150, and 229\,GHz Atacama Cosmology Telescope (ACT) observations that mapped ${\sim}40$ per cent of the sky for most of the data volume. Our data cover 88 gamma-ray bursts (GRBs), 12 tidal disruption events (TDEs) and 203 other transients, including supernovae (SNe). We stack our ACT observations to increase the signal-to-noise ratio of the maps. In all cases but one, we do not detect these transients in the ACT data. The single candidate detection (event AT2019ppm), seen at ${\sim}5\sigma$ significance in our data, appears to be due to active galactic nuclei (AGN) activity in the host galaxy coincident with a transient alert. For each source in our search we provide flux upper limits. For example, the medians for the 95 per cent confidence upper limits at 98\,GHz are $15$, $18$, and $16$\,mJy for GRBs, SNe, and TDEs respectively, in the first month after discovery. The projected sensitivity of future wide-area cosmic microwave background (CMB) surveys should be sufficient to detect many of these events using the methods described in this paper.
\end{abstract}



%%%%%%%%%%%%%%%%%%%%%%%%%%%%%%%%%%%%%%%%%%%%%%%%%%

%%%%%%%%%%%%%%%%% BODY OF PAPER %%%%%%%%%%%%%%%%%%

\section{Introduction} \label{sec:intro}

For more than a decade, ground-based cosmic microwave background (CMB) telescopes have been scanning large fractions of the sky---about 40 per cent in the case of the Atacama Cosmology Telescope (ACT)---at millimetre (mm) wavelengths with arcminute resolution. The next generation of CMB experiments, such as the Simons Observatory \citep[SO;][]{2019JCAP...02..056A} and CMB-S4 \citep{2019arXiv190704473A} will continue to observe large fractions of the sky with regular cadences of a few days and exceptional sensitivity. This is opening up the possibility for mm time domain science, a largely unexplored field. Recently, ACT serendipitously detected three events consistent with nearby stellar flares \citep{2021ApJ...915...14N}, while a systematic search of the South Pole Telescope (SPT) data detected 12 events that seemed to be stellar flares and two others that may be extragalactic in origin \citep{2021ApJ...916...98G}. Previously, SPT performed a blind search for transients, finding an extragalactic candidate at $2.6\sigma$ significance with a duration of a week \citep{2016ApJ...830..143W}.

% this paragraph introduces the general mechanism of synchrotron
Of particular interest are extragalactic transient sources dominated by synchrotron emission. In general, this emission originates when the fast-moving ejecta interacts with the circumstellar medium (CSM), accelerating free electrons to relativistic speeds, and emitting synchrotron radiation as a consequence. Among these transient sources are gamma-ray bursts (GRBs), tidal disruption events (TDEs) and supernovae (SNe). This work will focus on these three kinds of objects. While mm detections of individual objects exist \citep[see Table 1 of][for a comprehensive catalog]{2022ApJ...935...16E}, we do not have reliable predictions of their mm luminosities in many cases. Nevertheless, this region of the spectrum is important: many mm transients peak at early times after trigger, and certain emission components are best observed at mm wavelegths, such as the reverse shocks in GRB afterglows \citep{2018ApJ...862...94L}. 
Whereas follow-up observations of transients discovered by other facilities rely on assumptions about which transients will produce bright mm emission, a CMB mm survey instrument like ACT could provide a systematic measurement of mm emission from different classes of transients due to its large sky coverage and cadence, and may sometimes observe events at or shortly after the trigger, in contrast to some follow-up targeted observations that can take $\sim$hours to begin.

% This is the SNe paragraph
In the case of SNe, we focus on the core-collapse classes. They can produce bright radio emission arising in the interaction of the rapid ejecta with the CSM, generating synchrotron due to acceleration of electrons \citep{2002ARA&A..40..387W}. Despite this, only $\mathcal{O}(100)$ SNe have been detected in radio emission \citep[e.g.][]{1987Natur.327...38T,2002ApJ...581..404B,2021ApJ...908...75B}. Only a handful of mm observations of nearby SNe exist, such as SN2011dh \citep{2013MNRAS.436.1258H}, iPTF13bvn \citep{2013ApJ...775L...7C}, and SN2020oi \citep{2021ApJ...918...34M}. An interesting case is interacting SNe, a class of objects defined by their interaction with the CSM, characterised by the production of signature narrow spectral lines due to shocks \citep{2017hsn..book..403S}. Interacting SNe include the IIn and Ibn classes, which have been detected in radio \citep[e.g.][]{2018SSRv..214...27C}, but not in the mm so far. The progenitors for this class are not clearly identified. These classes of SNe might produce mm emission due to (sub)relativistic ejecta interacting with very dense ($n_e \gtrsim 10^6$\,cm$^{-3}$) CSM \citep{2022ApJ...934....5Y}. Since type IIn SNe take much longer than other core-collapse classes to become detectable in the radio \citep[their luminosity-rise time is $\gtrsim 1$ order of magnitude higher,][]{2021ApJ...908...75B}, a detection in the mm would probe the more extreme physical conditions closer to the explosion time and therefore help identify potential progenitors, due to the mm peak happening at much earlier times than the radio. The best example of a SN-like event that could have been easily observed by a CMB experiment was the Fast Blue Optical Transient (FBOT) AT2018cow \citep{2019ApJ...871...73H}, a bright and rapidly evolving transient that lasted ${\sim}100$ days, with a ${\sim}100$\,GHz flux of ${\sim}90$\,mJy 22 days after discovery. However, events like these are rare \citep{2023ApJ...949..120H}. As a reference, this event would have to be at a distance of 80 (130)\,Mpc to be detected at $3\sigma$ by the 98 (95)\,GHz channel of ACT (SPT) in a single visit. A relatively nearby supernova, like SN~2011dh, at a distance of only 7.5\,Mpc, has a measured flux of ${\sim}4$\,mJy at 107\,GHz four days after discovery that decays thereafter \citep{2013MNRAS.436.1258H}. While we expect core-collapse SNe to produce potentially detectable mm emission, we will also report on observations of type Ia SNe or unconfirmed events.

% This is the TDE paragraph
TDEs correspond to the destruction of stars by the tidal forces of supermassive black holes at galaxy centres. The synchrotron radiation originates from the interaction between the TDE outflows and the surrounding circumnuclear medium (CNM). This interaction drives a bow shock into the gas, which accelerates electrons to relativistic speeds and produces synchrotron emission. However, we can expect a wide variety of emission from event to event due to dependence on multiple factors, such as viewing angle, how much ambient gas there is, fraction of shock energy deposited in electrons and magnetic field, etc. \citep{2020SSRv..216..114R}. Due to this, there is a large diversity of radio properties in the TDEs for which we have radio and mm detections \citep{2020SSRv..216...81A}. In broad terms, they can be classified into radio loud and radio quiet (with the frontier between the two being a radio characteristic luminosity $\sim 10^{40}$\,erg\,s$^{-1}=2.6 \times 10^6\,L_{\sun}$). Radio loud TDEs are very bright due to their relativistic jets typically being viewed on-axis and their flux Doppler boosted, while radio quiet TDEs produce non-relativistic outflows that have much weaker mm emission than radio loud TDEs, or none at all, although this latter case might also be consistent with a relativistically jetted TDE being viewed off-axis. Both types of TDE have been observed in the radio, but only a handful of radio quiet detections exist, which may be an observational bias \citep{2020SSRv..216...81A}. Due to the small number of TDEs discovered to date, it is not clear what fraction of them will produce high-energy jets or which fraction will produce radio emission. However, the detection of TDE jets in the mm probes the CNM density and magnetic field around supermassive black holes on sub-parsec scales \citep{2016MNRAS.461.3375Y}. The best and most extensively studied examples of TDEs in radio/mm emission are the radio loud and very bright Sw J1644+57 \citep{2011Natur.476..425Z,2012ApJ...748...36B} and  the radio quiet ASASSN-14li \citep{2016Sci...351...62V,2016ApJ...819L..25A}. \citet{2021ApJ...919..127C} reported Very Large Array (VLA) and Atacama Large Millimeter Array (ALMA) observations of the non-relativistic TDE AT2019dsg. They measure a 100\,GHz flux of 0.07\,mJy 74 days after discovery with ALMA. Also, \citet{2016MNRAS.461.3375Y} found the TDE IGR~J12580+0134 in archival \Planck data, likely due to non-relativistic emission from an off-axis jet.

% GRB paragraph
Long GRBs (with a duration longer than $\sim 2$\,s) have their origin in highly energetic explosions produced by the core collapse of massive stars in high-redshift galaxies. This explosion creates the ejection of material at ultra-relativistic speeds through highly collimated jets \citep{2003Natur.423..847H}. As the jet penetrates into the surrounding gas, a forward shock expands outwards into the gas, producing a cascade of emission at wavelengths longer than gamma rays, known as the GRB afterglow. A reverse shock into the jet ejecta produces additional emission \citep{1999PhR...314..575P}. The accelerated electrons in this interaction will generate a synchrotron spectrum with several break frequencies and a defined power-law behaviour. This simplified model is known as the fireball model \citep{1998ApJ...497L..17S}, which predicts that the afterglow light curve will peak at earlier times in the mm than in the radio. The forward shock can peak on scales of several hours to a few days in the $\sim 80{-}400$\,GHz region of the spectrum \citep{2012A&A...538A..44D}, while the reverse shock can peak on much shorter timescales of only half to a few hours \citep{2018ApJ...862...94L,2019ApJ...878L..26L}. The advantage of detecting mm emission from GRBs afterglows is that it allows synchrotron emission to be studied during the first few days of the event, allowing many or even all the parameters that model the fireball can be constrained. The mm is also in the sweet spot of not being affected by self-absorption at lower frequencies or by the dust extinction at higher frequencies. A typical long GRB at $z \sim 1$ ($ \sim 6.7$\,Gpc) is expected to peak at ${\sim}1{-}2$\,mJy a few days after the burst at 100\,GHz \citep{2022ApJ...935...16E}. All of the mm detections for GRBs until 2012 can be found in \cite{2012A&A...538A..44D}, while an updated list can be found in \cite{2022ApJ...935...16E}. An example of a nearby GRB extensively studied in the mm is GRB 130427A \citep[e.g.][]{2013ApJ...776..119L,2014ApJ...781...37P}.

In this paper, we search for SNe, TDEs, GRBs, and assorted astronomical transients (ATs)\footnote{AT is one of the official classification by the International Astronomical Union (IAU) for reported transients discovered by the astronomical community. Once spectroscopically confirmed, a supernova will receive its SN name. Most of the ATs are SN candidates that were not followed up or classified. The others are  assorted astrophysical phenomena.} discovered at other wavelengths in ACT mm-wave observations between 2013 and 2021. To give a sense of the detection prospects, the ACT 98\,GHz frequency band is capable of detecting a 50\,mJy point source at $3\sigma$ with a single observation of two detector arrays (see Section~\ref{ssec:ACT} for the definition of an array).

While ACT lacks the sensitivity to detect the typical mm flux expected from the types of sources we are targeting, there is always the possibility of discovering unexpectedly high emission from an unusual event, such as AT2018cow,\footnote{Note that this event is not in our search space: its declination was slightly above ACT's survey footprint, and it also occurred during a period when the telescope was not observing.}  and we can also probe to lower fluxes by stacking over longer time spans. Furthermore, it is now opportune to use the many years of ACT data to develop techniques for performing systematic searches for transient events that will further increase the science output of future experiments, like SO and CMB-S4, which will have better sensitivity.

This paper is organised as follows: Section~\ref{sec:observations} describes the ACT observations and the matched filter method used to measure the flux from the maps. Section~\ref{sec:selection} describes how we choose the sources to cross reference with the ACT observations. In Section~\ref{sec:results}, we show our results. Finally, we present our conclusions in Section~\ref{sec:conclusions}.

For calculating cosmological distances, we assume a flat cosmology with $\Omega_{\rm m}=0.31$ and $H_0=67.7$\,km\,s$^{-1}$\,Mpc$^{-1}$, as measured by \Planck Data Release 3 \citep{2020A&A...641A...6P}.
\section{Observations and Methods}\label{sec:observations}

\subsection{ACT}\label{ssec:ACT}

In this paper, we use observations from the second and third generations of receivers, ACTPol \citep{2010SPIE.7741E..1SN,2016ApJS..227...21T} and AdvACT \citep{2016JLTP..184..772H,2018JLTP..193..267C}, respectively. These used the same cryostat, which houses three optics tubes, each containing an array of transition-edge sensor (TES) bolometers. Arrays were occasionally changed between observing seasons, such that the data analysed in this paper come from six arrays, denominated PA1 through PA6. Together, these cover three bandpasses: f090 (77--112\,GHz), f150 (124--172\,GHz), and f220 (182--277\,GHz).\footnote{These frequency ranges contain 99 per cent of the power in the bandpass.} ACTPol began with PA1 (2013--2015) and PA2 (2014--2016), which observed in the f150 band. The dichroic PA3 (2015--2016) array observed at both f090 and f150. AdvACT replaced PA1 with PA4 (2016 onward) which observed both the f150 and f220 bands, and then replaced PA2/3 with PA5 (2017 onward) and PA6 (2017--2019), each of which observed in both the f090 and f150 bands.\footnote{In 2020, PA6 was replaced by PA7 containing the low frequency channels f030 and f040, but these channels are not included in our study since our analysis of their data is not yet mature enough.} Table~\ref{tab:freq_channel} summarises the frequency channel definitions used in this paper, the effective band centre and band width of each channel and array assuming a synchrotron spectrum $S(\nu) \propto \nu^{-0.7}$, consistent with synchrotron-dominated extragalactic sources, and the dates of the data we searched within.

\begin{table*}
    \caption{
    Frequency channels and dates of the data included in the search. Note that there are gaps up to $\mathcal{O}$(months) within the time ranges indicated, due to climate conditions, yearly planned maintenance, upgrades, etc.
    \label{tab:freq_channel}
    }
    \begin{tabular}{ c c c c c c c }
            \hline
            Channel & Array & Band & Bandwidth & Synchrotron Band & Data Start Date & Data End Date \\
            & & Centre (GHz)$^{a}$&(GHz)$^{b}$& Centre (GHz)$^{c}$&& \\
            \hline
            \hline
            f090 & PA3 & 93.3 & 31.1 & 93.2 & 2015 April 21 & 2016 December 22 \\
             & PA5 & 96.5 & 19.0 & 96.5 & 2017 May 11 & 2021 June 18\\
             & PA6 & 95.3 & 23.1 & 95.3 & 2017 May 11 & 2019 December 19\\
            \hline
            f150 & PA1 & 145.4 & 39.6 & 145.3 & 2013 September 10 & 2016 June 12 \\
             & PA2 & 145.9 & 36.7 & 145.7 & 2014 August 23 & 2016 December 22 \\
             & PA3 & 144.9 & 27.8 & 144.7 & 2015 April 21 & 2016 December 22 \\
             & PA4 & 148.5 & 36.7 & 148.3 & 2017 May 11 & 2021 June 18 \\
             & PA5 & 149.3 & 28.1 & 149.2 & 2017 May 11 & 2021 June 18 \\
             & PA6 & 147.9 & 31.1 & 147.8 & 2017 May 11 & 2019 December 16 \\
            \hline
            f220 & PA4 & 226.7 & 66.6 & 225.0 & 2017 May 11 & 2021 June 18 \\
            \hline
            \multicolumn{7}{l}{\footnotesize $^a$The effective frequency is defined as $\nu_0=\frac{\int \nu \tau(\nu) d\nu}{\int \tau(\nu) d\nu}$, where $\nu$ is the frequency and $\tau(\nu)$ is the passband of the channel as a function of frequency.} \\
            \multicolumn{7}{l}{\footnotesize $^b$The bandwidth is defined as $\int \tau(\nu) d\nu$.} \\
            \multicolumn{7}{l}{\footnotesize $^c$For a $\nu^{-0.7}$ spectrum and assuming a point source.} \\
    \end{tabular}
\end{table*}
We only use data taken during the night, defined as falling between 23:00 and 11:00 Coordinated Universal Time, since the telescope has consistent and stable beam profiles during these times.\footnote{Better characterising the day time data so that it can be used for science is an active area of research within the collaboration \citep[e.g.,][]{2020JCAP...12..046N,2021ApJS..253....3H}.} We make maps in specific time intervals (detailed in Section~\ref{sec:selection}) using the standard ACT Maximum Likelihood (ML) mapmaker \citep{2020JCAP...12..047A} inside a $2\times2\,\mathrm{deg}^2$ stamp centred on the coordinates of the respective transient, using 10 conjugate gradient steps---enough for the map to converge on the small angular scales relevant for compact sources---and with a downsampling factor of two in the datastream to speed up the map-making process. A given map will include one or more individual observations by the telescope that are in the corresponding time range. An individual observation is a $\sim$10-minute constant-elevation scan.

Our matched filter (see Section~\ref{sec:matched-filter}) takes the instrumental beam as input, for which we use an empirically measured beam determined for each combination of detector array and frequency channel, as described in \citet{2020JCAP...12..047A}. To briefly summarise the process, nighttime maps of Uranus are azimuthally averaged to obtain the radial beam profile and the harmonic beam window function. We use the `jitter' beams \citep{2022JCAP...05..044L}, which include corrections to account for small pointing variations from night to night and are calculated on maps after additional corrections take place due to pointing jitter. They also made the small correction for the different response across each frequency passband to the CMB blackbody spectrum and a Rayleigh-Jeans spectrum expected for Uranus. Finally, we apply a first-order correction to the harmonic transform of the beam $B(\ell)' = B(\ell \nu_{\rm CMB}/\nu_{\rm syn})$ \citep{2013ApJS..209...17H}, where $\nu_{\rm CMB}$ is the effective band centre of the channel and array for a CMB blackbody spectrum and $\nu_{\rm syn}$ is the effective band centres for a $S(\nu) \propto \nu^{-0.7}$ synchrotron spectrum listed in Table~\ref{tab:freq_channel}.

\subsection{Matched filter} \label{sec:matched-filter} To measure the flux of sources in our maps, we use the matched filter (MF) approach, which maximizes the signal-to-noise ratio (S/N) of point sources by inverse covariance-filtering the beam-deconvolved map. We follow the approach derived in detail in Section\,4.3.2 of \cite{2021ApJ...923..224N}; below we summarize the process.

In terms of the standard (non-deconvolved) brightness-temperature map $\vect{m_{\rm T}}$, the flux is given by
\begin{equation} \label{eq:rho}
    \vect{\rho} = \matr{B} \matr{N}^{-1} \vect{m_{\rm T}} \text{,}
\end{equation}
where $\matr{B}$ is the instrument beam (diagonal in harmonic space) and $\matr{N}$ is the noise covariance matrix of the map (which for the purpose of point source finding includes not only instrument and atmospheric noise, but also the CMB and other foregrounds/backgrounds). Equation~\ref{eq:rho} applies in both the real and the harmonic basis. This definition maximizes S/N, while the following normalization factor $\kappa$ is useful to compute the physical units required for the analysis:
\begin{equation} \label{eq:kappa}
    \vect{\kappa} = \mathrm{diag}( \matr{B} \matr{N}^{-1} \matr{B}) \text{.}
\end{equation}
With this we can calculate the flux map, $\vect{f}$, its standard deviation, $\Delta \vect{f}$, and the S/N map, $S/N$, as:\footnote{$\vect{f}$ can be interpreted as the maximum-likelihood estimate for the flux in each pixel under the assumption that all flux comes from a point source at the centre of that pixel. It is optimal and unbiased as long as point sources do not get closer than one beam to each other.}
\begin{align} \label{eq:flux-dflux}
    \vect{f} &= \vect{\rho}/\vect{\kappa}, & \Delta \vect{f} &= \vect{\kappa}^{-\frac12}, & S/N = \vect{\rho} \vect{\kappa}^{-\frac12} \text{.}
\end{align}
Note that $\vect{m_{\rm T}}$, $\vect{\rho}$ and $\vect{\kappa}$ are all vectors in real space (with each element representing a pixel in the map), so the divisions here are done element-wise. ACT maps normally represent the variations, in $\si{\micro\kelvin}$, around the mean CMB blackbody temperature, but we transform them to units of Jy\,sr$^{-1}$, by multiplying by the derivative of the blackbody spectrum $I_{\rm BB}$ with respect to the temperature, $\partial I_{\rm BB}/\partial T \times 10^{-6}$, evaluated at the frequency of the corresponding channel and $T_{\rm CMB}=2.725$\si{\kelvin} \citep{2009ApJ...707..916F}. The factor of $10^{-6}$ accounts for the \si{\micro \kelvin} to \si{\kelvin} conversion.

As a ground-based microwave telescope, ACT has to look through the spatially and temporally varying water vapour distribution of the atmosphere, whose turbulence acts as a large source of correlated noise. This leads to the map noise covariance matrix being complicated to model. We use the estimate $\matr{N}^{-1} = \matr{H} \matr{C}^{-1} \matr{H}$, where $\matr{H} = \matr{W}^{\frac12}$, and $\matr{W}$ is the inverse variance of $\vect{m_{\rm T}}$, a matrix representing an estimate of the white noise inverse variance per pixel, which is available as an output from the mapmaker. We employ the Fourier-diagonal approximation $\matr{C}^{-1} = 1/(1+(\vect{k}/k_{\rm knee})^{-3})$ for the noise correlation properties,\footnote{We tested using the empirical power spectrum directly calculated from the map, masking the central area, instead of using this approximation for the noise spectrum. The approximation recovers a consistent flux, while the error bar is ${\sim}3{-}5$ per cent higher. For simplicity of calculations, we use the approximation for the $1/f$ noise spectrum.} with $k_{\rm knee} = 2000$ for f090, 3000 for f150, and 4000 for f220.\footnote{This could be improved by also modelling the directionality of the noise correlations, which could improve the S/N by ${\sim}$10--20 per cent (possibly more in very stripy regions).} As such, the noise covariance matrix, $\matr{N}$, contains correlated $1/f$ noise modulated spatially by the inverse covariance $\matr{W}$. Note that $\matr{B}$, $\matr{H}$, $\matr{C}$, and $\matr{N}$ are matrices; $\matr{B}$ and $\matr{C}$ are harmonic representations and are diagonal in harmonic space (since we treat the beam as azimuthally symmetric), while $\matr{H}$ is a pixel representation and diagonal in pixel space (we do not compute the correlations among pixels), as is $\matr{W}$.

\begin{figure}
    \centering
    \includegraphics[width=0.9\columnwidth]{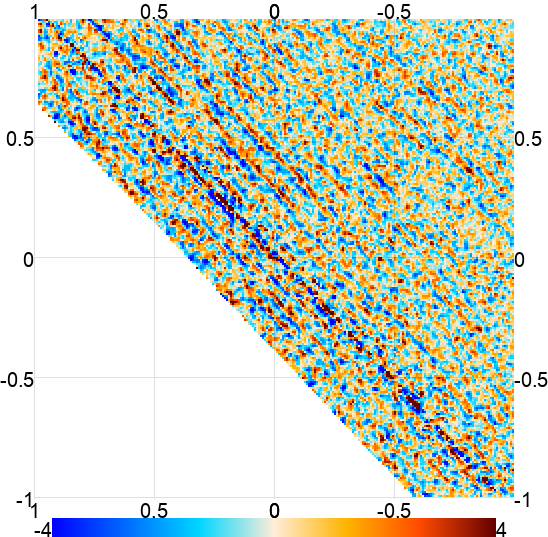}
    \caption{Map of the flux S/N of an individual map centred on one of our GRB sources, exemplifying the stripy noise that often occurs when there is poor coverage near the edge of an observation. The range is $\mathrm{S/N} = \pm 4$.}
    \label{fig:example_stripy}
\end{figure}

\subsection{Excising Noisy Maps} \label{sec:noisy-maps}

%paragraph about stripy noise and how we discard them

The telescope's scanning pattern results in stripy noise in the maps, which is not captured by our isotropic noise model (see section~\ref{sec:matched-filter}). For most of our maps this produces a negligible effect, but a small fraction of the maps are stripy enough that it becomes a problem, in two different ways:
\begin{enumerate}
    \item Mismodelled noise is not properly weighted when averaging, leading to an overall loss in S/N.
    \item An incorrect noise model leads to incorrect error bars in the final measurements. The net result is that noise fluctuations are misinterpreted as signal.
\end{enumerate}
The optimal way to handle this would be to generalise the noise model to support anisotropic noise,\footnote{This could be done by measuring the 2D noise power spectrum near the sources, and then using this as $\matr{C}$ in the matched filter.} but for now we simply identify and cut maps that satisfy any of the following criteria.

\begin{enumerate}
    \item The number of sample hits per pixel at the centre of the map is $\leq 25$ hits.
    \item The hits map in a $5 \times 5$\,arcmin$^2$ stamp around the centre has more than 30 percent of its pixels with $\leq 25$ hits.
    \item The hits map in a $20 \times 20$\,arcmin$^2$ stamp around the centre has more than 50 percent of its pixels with $\leq 25$ hits.
\end{enumerate}
This results in about 6\% of the maps being cut. While some visible striping is left after this cut, it is no longer at the level where it significantly affects our results. Appendix~\ref{sec:stripy-noise} gives an example of how our excision of stripy maps improves the accuracy of our measurements.

\subsection{Astrophysical background} 

For every matched filter map we produce, we account for the astrophysical background, which often might include the host galaxy of the particular transient plus any other contamination. We do this using depth-1 maps.\footnote{Depth-1 maps are high-resolution ACT maps made from a single contiguous constant-elevation scan lasting less than 24 hours, meaning the telescope only drifted past each point in the map once. This makes the time at which each pixel was observed unambiguous, allowing for $\sim$minute precision timestamps for events. Each pixel in a map contains about 4 minutes of integration time, which is roughly how long it takes for a given coordinate in the sky to cross a detector array.} 
To account for any static, background flux for a transient, we stack all the depth-1 maps from a 350-day period before and after the time window we consider for the transient event (see Section~\ref{sec:selection} for details). We then match filter this coadded map to estimate the background flux at the coordinate of the transient, and subtract it from the fluxes of the individual maps during the time window of the transient event.
However, there is a possibility of variable background due to galactic and/or AGN activity from the host, or even nearby sources that leak into the beam.

To estimate this contamination, we split the two 350-day periods into two coadded maps, one before and one after the transient event. We measure the difference in flux between the two maps, and divide this difference by the error bar (which is the square root of the sum of the error bars of each of the two maps). We perform this for all maps and sources for which we have such measurements both before and after the transient event. If these measurements are taken from sources whose flux does not vary in time, then any difference between the two 350-day periods should be only due to statistical fluctuations consistent with the error bars, and therefore the sample should be consistent with a normal distribution with zero mean and unit standard deviation. According to the Anderson-Darling test \citep[e.g.,][]{doi:10.1080/01621459.1974.10480196}, this sample is consistent with normality. The test statistic is 0.46, which cannot reject the null hypothesis of normality at 85\% significance level.
As a sanity check to verify that this procedure is sensitive to variability, we repeated the same exercise of calculating the difference of flux in the stack of two distinct years (2018 and 2019) for 206 of the highest S/N AGNs in the ACT field, the majority of which are highly variable. (A separate paper on AGN light curves is in preparation.) Using the same normality test, the null hypothesis of zero variability is rejected at more than 99\% significance, as expected.
Nevertheless, there are outliers in our sample of flux differences between the two 350-day stacks. The largest outlier has a difference at the $-4.8\sigma$ level and exhibits flare activity in the host galaxy after the transient discovery, while it is quiet before. This host galaxy shows evidence of relatively short flares in previous years, while in the full survey maps it is not detected. In conclusion, while we do expect some of our sources to include variable backgrounds, these outliers are not enough to break the normality of the flux difference sample at our current level of sensitivity.

We have only produced depth-1 maps from the 2017 season onward; where the depth-1 maps are lacking, we use the Data Release 5 (DR5) coadded ACT maps at the relevant frequency from \cite{2020JCAP...12..046N} (which are coadds of 11 ACT seasons in the period 2008--18) to estimate the astrophysical background. In all cases, we mask bright point sources from the ACT catalogue with flux above 100\,mJy to a radius of 3\,arcmin.\footnote{For the SN candidates, we use 5\,arcmin instead since these sources are usually in the local Universe and nearby, very bright, extended galaxies need to be removed.}

Since we are subtracting two flux matched filter maps---the short-time scale map, $\vect{f}$, with inverse covariance, $\vect{\kappa}$ (equations~\ref{eq:flux-dflux} and \ref{eq:kappa}, respectively), and the background map, $\vect{f}_{\rm bkg}$, with inverse covariance $\vect{\kappa}_{\rm bkg}$---we need to calculate the inverse covariance of the new map.\footnote{Subtracting filtered maps like this is not exact, because filters differ slightly. Ideally the subtraction would happen before filtering. However, we find this approximation to be good enough in practice.} If $\vect{f}' = \vect{f} - \vect{f}_{\rm bkg}$, then by equation~\ref{eq:flux-dflux}, $\vect{\kappa}' = \vect{\kappa}\vect{\kappa}_{\rm bkg}/(\vect{\kappa}+\vect{\kappa}_{\rm bkg})$.

In some cases, our maps of transient candidates contain only one or a few individual observations, typically because coverage of that part of the sky is poor due to being near the edge of the field of observation. We do not report a flux when its error bar is $>500$\,mJy. Furthermore, to account for uncertainty due to variations in the flux calibration from observation to observation, we increase the measured flux error bars using the estimated standard deviation of fractional residual light curves from Uranus observations. For f090, this uncertainty is ${\sim}$1{--}3 per cent, for f150 it is ${\sim}$2--8 per cent, and for f220 it is ${\sim}$4{--}12 per cent. Appendix~\ref{sec:calibration} describes this measurement and procedure in detail.
\section{Selection of extragalactic sources} \label{sec:selection}

For our targeted extragalactic transient search, we choose GRBs detected with the Neil Gehrels \Swift Observatory \citep{2004ApJ...611.1005G}, SNe and ATs from the comprehensive Open Supernova Catalog \citep{2017ApJ...835...64G}, and the TDEs listed in the recent review by \citet{2021ARA&A..59...21G} as well as TDEs recently discovered in X-rays by the SRG All-Sky Survey \citep{2021MNRAS.508.3820S} at the eROSITA instrument aboard the Spektr-RG space observatory \citep{2021A&A...647A...1P,2021A&A...656A.132S}. We consider multiple time ranges for the maps we produce, all measured relative to the discovery date. These are in ranges of 3 consecutive days, 7 consecutive days (1 week), 28 consecutive days (1 month), 56 consecutive days (2 months), and 84 consecutive days (3 months). We make maps for the sources for which we have ACT observations in at least one of the epochs above. The exact time ranges are detailed below. In total, we make at least one map for 203 distinct SNe/ATs, 12 distinct TDEs, and 88 distinct GRBs.

\begin{figure*}
    \centering
    \includegraphics[width=0.33\textwidth]{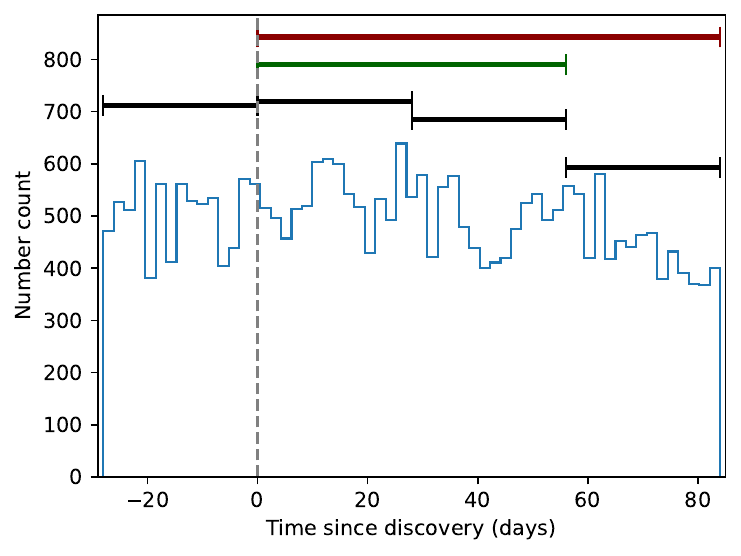}
    \includegraphics[width=0.33\textwidth]{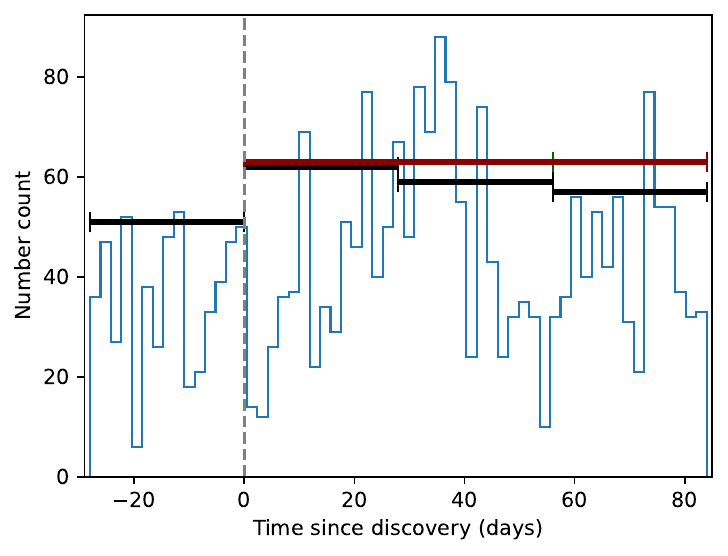}
    \includegraphics[width=0.33\textwidth]{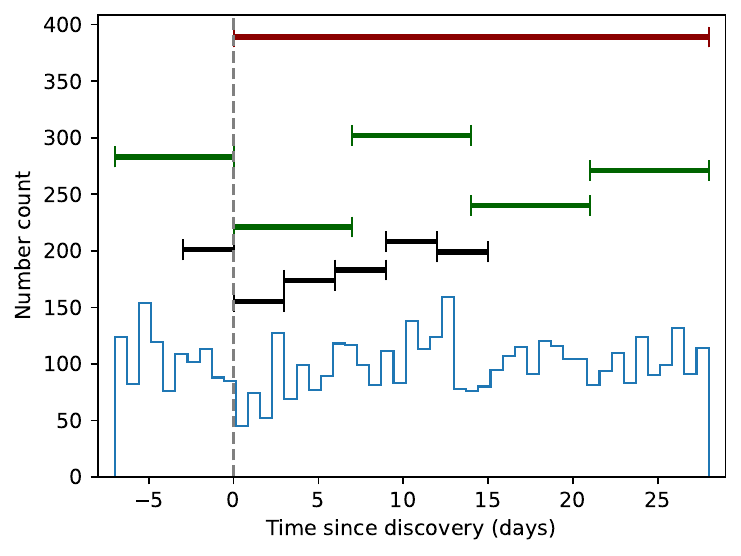}
    \caption{Number of maps per time range relative to transient discovery time. The horizontal bars correspond to the number of maps we have for each epoch we consider for each transient type (SNe and ATs to the left, TDEs in the centre, and GRBs to the right). Their widths show the time range of the epoch. The different colours correspond to different time lengths of the interval, described in the main text. Since we make maps per frequency channel and array, one source might be observed multiple times in a given epoch. The thin blue line in each panel is the histogram of individual ACT observations that scan the sources we target.
    }
    \label{fig:epochs}
\end{figure*}

Our observations are distributed approximately uniformly in time because ACT operates as a survey instrument, performing wide scans in azimuth at fixed elevations while the sky rotates through its field of view. Thus, if a given coordinate is contained in one of ACT's survey areas, it will be visited with a fairly uniform cadence over periods of 1--3 months. Figure~\ref{fig:epochs} shows a histogram of the number of individual ACT observations that included one or more sources from our selection, measured relative to the time of discovery. The distribution is relatively flat due to the survey observing strategy (though in the case of TDEs the distribution is noisier because there are only 12 sources of this type). Figure~\ref{fig:epochs} also shows with horizontal bars the time ranges that we bin in (i.e., the 3, 7, 28 etc. days described above) along with the total number of maps included in each range.

\begin{figure}
    \centering
    \includegraphics[width=1.0\columnwidth]{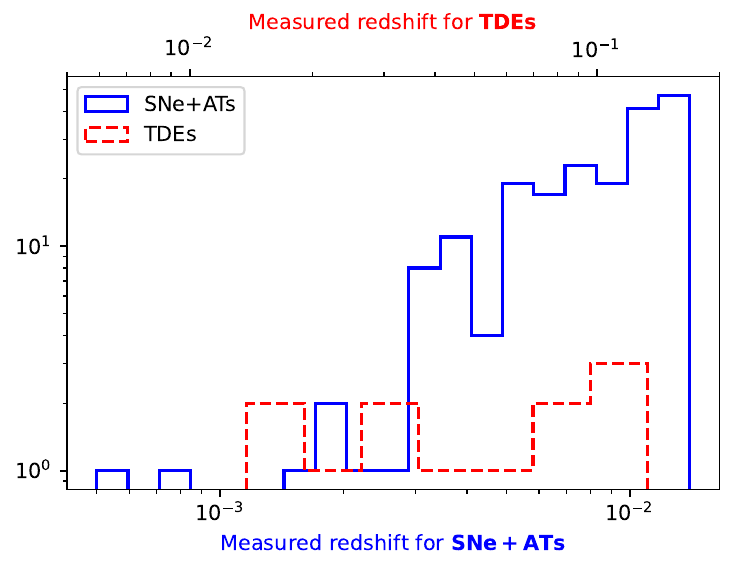}
    \caption{
    Histograms of the measured redshifts for the 203 SNe+ATs and 12 TDEs observed by ACT and that we include in our analysis. For the 88 GRBs we observe, only $\sim 27$\% have a measured redshift, with values $0.14 \lesssim z \le 3.5$.
    }
    \label{fig:properties}
\end{figure}

\subsection{SNe and ATs} We select transients from the Open Supernova Catalog that compiles SNe discovered from multiple sources and multiple surveys. Since it also includes reported ATs, we include them in this category. Also, a few cases were solar system objects mistaken for SNe. We checked all of the sources we include in our SN/AT catalogue in the Transient Name Server\footnote{https://www.wis-tns.org} to make sure they are real extragalactic sources and removed the rest. Although there are several thousand sources in the time range of our observations, in reality a SNe must be relatively nearby to be detectable by contemporary mm-wave telescopes. Since FBOT AT2018cow is an example of an uncommon SN-like event that could be detected with a high S/N by mm experiments \citep{2019ApJ...871...73H,2019ApJ...878L..25H}, we use its redshift $z=0.014$ \citep{2019MNRAS.484.1031P} as our upper limit for SNe candidates. For the few examples of observed SNe in mm bands, the emission is months long, motivating us to produce maps on month-long intervals, starting a month before the discovery date and continuing to the third month after discovery (black bars in Figure~\ref{fig:epochs}, left). Additionally, we make two- (dark green bars) and three-month maps (dark red bars) after the discovery. In Figure~\ref{fig:properties}, we show the distribution of measured redshifts for the population, which is in the range $z=0.00001-0.07$.

\subsection{TDEs} There are fewer than 100 discovered TDEs in the literature. We search for those listed in the review by \citet{2021ARA&A..59...21G}, which includes all TDEs identified until 2019. We also use the sample of TDEs discovered during 2020 in X-rays by the SRG All Sky Survey made with the eROSITA instrument \citep{2021MNRAS.508.3820S}, although of the 13 in their catalogue, only two are at low enough declination to appear in the ACT survey region. Since we have examples of TDEs lasting for several months, we use the same time intervals for maps that we use for SNe (see Figure~\ref{fig:epochs}, centre, which uses the same colour scheme as in the previous subsection). In Figure~\ref{fig:properties}, we show the distribution of measured redshifts for the population, which is in the range $z=0.0151-0.132$. In our sample of observed TDEs, there are no known jetted events. Two of them are events with detected radio emission and non-relativistic outflows, while three of them have no detected radio emission. The rest are either possible TDEs, SRG All Sky Survey or Zwicky Transient Facility-discovered sources \citep{2021ApJ...908....4V} that have not been studied in detail in radio/mm.

\subsection{GRBs} \label{ssec:selection_grbs} GRBs are selected from the database of the \Swift observatory. The mission discovers ${\sim}100$ GRBs per year, using the Burst Alert Telescope (BAT), which has a wide field of view and operates in the gamma ray range of 15--150\,keV. Within 90 seconds of a trigger from BAT, the satellite is pointed to the approximate position of the source and observes it with the X-Ray Telescope (XRT) and UV/Optical Telescope (UVOT), obtaining precise coordinates for each event. We use the coordinates of the GRB as measured by the XRT, which have a precision of a few arcseconds. The GRB afterglow emission is visible for hours and even several days in the mm range. For this reason, we produce maps of the observations in blocks of three days (black bars in Figure~\ref{fig:epochs}, right), starting three days before and up to 15 days after the discovery of the GRB. We also produce maps with seven days of observations, starting seven days before and finishing 28 days after the discovery of the GRB (dark green bars). Finally, we produce a one month map using the 28 days after the discovery day of the GRB (dark red bars). The redshift range for the GRBs in our observations is $z=0.1475-3.503$.

\section{Results} \label{sec:results}

For every time interval before and after the discovery date of a transient specified in Section~\ref{sec:selection}, we produce a map for each detector array and frequency channel that was observing at the time.
This results in a total of 6259 maps used in our analysis, after excising 410 noisy maps according to the prescription of Section~\ref{sec:noisy-maps}.
For each individual map, we estimate the excess flux over the background at the location of the transient candidate with the procedures described in Section~\ref{sec:matched-filter}. 

For sources with measurements in the same frequency channel and time interval, but from two or more arrays, we calculate the mean flux $\hat{f}$ by weighting each map with its inverse variance at the position of the transient source. That is, $\hat{f} = \sum_i f_i \Delta f_i^{-2} / \sum_i \Delta f_i^{-2}$ and $\Delta \hat{f} = (\sum_i \Delta f_i^{-2})^{-1/2}$, where $f_i\pm\Delta f_i$ is the flux measurement at the transient position for array $i$.

As we do not detect mm counterparts with high confidence for any of the transients in our search, except in the case of AT2019ppm, described below, all of our results are upper limits, quoted at the 95 per cent confidence level. We calculate this by integrating the Gaussian probability density function, $p(f)$, which has the mean of the measured flux, $f$, and a standard deviation equivalent to the error bar from the map (equation~\ref{eq:flux-dflux}), to obtain $f_{95}$, implicitly defined by the equation:
\begin{equation}
    \int_{0}^{f_{95}} p(f) df = 0.95  \int_{0}^{\infty} p(f) df.
\end{equation}
Note that we discard the negative portion of the Gaussian distribution.

\begin{figure}
    \centering
    \includegraphics[width=\columnwidth]{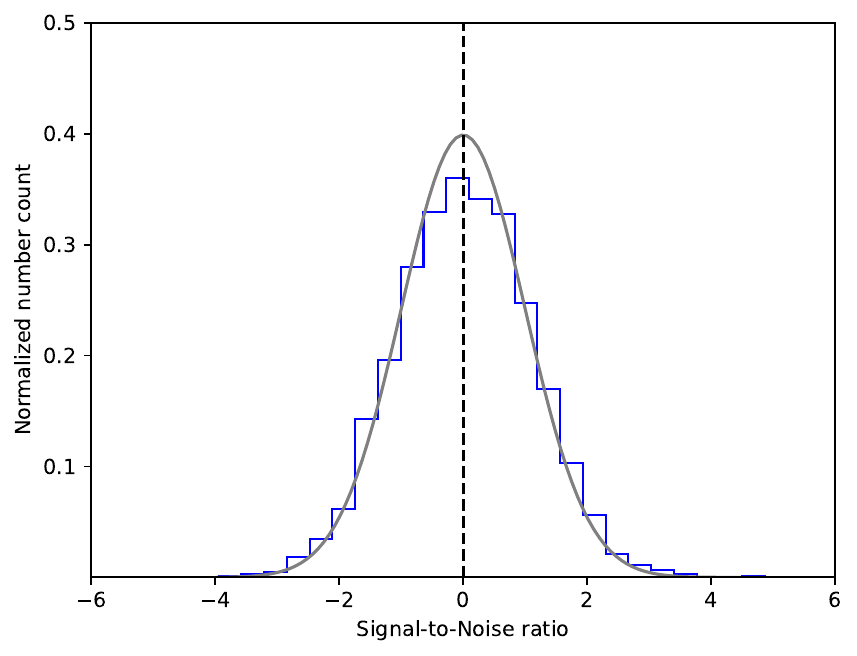}
    \caption{Histogram of the S/N of the flux measurements for all time intervals for all transients types in all frequency channels and all detector arrays. A standard normal probability density function is overplotted in grey.
    }
    \label{fig:histogram}
\end{figure}

The histogram of the S/N of our measured flux excess for the candidate transient sources is shown in Figure~\ref{fig:histogram}, with a standard normal probability density function overplotted. According to the Anderson-Darling test \citep[e.g.,][]{doi:10.1080/01621459.1974.10480196} it is consistent with a normal distribution, indicating that the ensemble of our events is consistent with random fluctuations in the map with no evidence for transient detections in the total sample of measurements. The test returns a value of 0.413, which fails to reject the hypothesis of normality at 85\% confidence. We obtain a similar result if we remove the AT2019ppm event (see Sec.~\ref{sec:AT2019ppm}, below) from the histogram. Applying the D'Agostino and Pearsons's skewness- and kurtosis-based omnibus test \citep{dagostino/pearson:1973, dagostino/belanger/dagostino:1990} also shows weak evidence of normality, with a $p-$value~$=0.031$. %, %which fails to reject the null hypothesis of normality at 95\% confidence. 
By removing the AT2019ppm event, we have $p-$value~$=0.165$, which is quite consistent with normality.

Investigating the higher S/N events in more detail, from our 6259 maps, we would expect 2 to 16 events (inclusive) to have S/N~$>3$ with 98.4 per cent confidence if they are normally distributed. We find 21 maps with S/N~$>3$. However, three of them correspond to the AT2019ppm event (see Sec.~\ref{sec:AT2019ppm}, below), and four of them  are clearly due to stripy noise features that were not detected by our hits map-based excision step (Section~\ref{sec:noisy-maps}). Discounting these seven, the number of S/N~$>3$ events is not unexpected; the fluxes in the centres of the 14 remaining maps appear consistent with random noise fluctuations. Two events have S/N~$>4.5$, which is unlikely (0.022 per cent) for our sample size. However, one of these (S/N $= 4.7$) is from AT2019ppm, while the other (S/N $= 4.9$) corresponds to a stripy noise map that was not rejected by our hits map algorithm.

Tables~\ref{table:sne}, \ref{table:tdes}, and \ref{table:grbs} show 95 per cent upper limits or measured fluxes for a few examples of SNe/ATs, TDEs, and GRBs. The columns show the upper limit in the corresponding time interval (in days, with zero being the discovery date of the event). The blank spaces correspond to time intervals where we do not have observations, or cases of longer time intervals that would be redundant (e.g., a map of the first seven days having the same information as a map of the first three days). The tables for the full list of transients that are investigated in this work are available as supplementary material as machine readable tables. As a reference, the median values for the 95 per cent upper limits at f090 are 18, 16, and 15\,mJy for the measurements for SNe, TDEs, and GRBs, respectively. 

% TABLES
\begin{table*}
\caption{Examples of the upper limits on flux density for SNe and ATs. The columns correspond to the time range of the map in days, with 0 being the discovery time. The numbers are 95 per cent upper limits and in mJy, except in the case of detections for AT2019ppm. The position is in degrees. \label{table:sne}}
\centering
    \begin{tabular}{  l  l  l  l  r  r  r  r  r  r  }
        \hline
        &  & & & \multicolumn{6}{c}{Time interval for the map relative to alert (in days)} \\
        \colheadd{Transient name} & \colheadd{RA/Dec} & \colheadd{Discovery} & \colheadd{Freq.} & \colheadd{[$-$28,0]} & \colheadd{[0,28]} & \colheadd{[28,56]} & \colheadd{[56,84]} & \colheadd{[0,56]} & \colheadd{[0,84]} \\
         & deg. & & & mJy & mJy & mJy & mJy & mJy & mJy \\
        \hline
        \hline
        \multirow{1}{*}{SN2013ft}&355.3967,3.7251&2013-09-13& f150&15.8&3.4&2.5&3.5&1.9&1.7\\
        \hline 
        \multirow{3}{*}{SN2017gax}&71.4560,-59.2452&2017-08-14& f090&15.8&17.1&11.0&12.5&9.8&7.9\\
        &&& f150&7.5&21.3&8.2&17.1&8.0&7.1\\
        &&& f220&67.0&55.1&61.6&110.0&43.0&41.8\\
        \hline
        \multirow{3}{*}{SN2018hdp}&33.4377,4.1023&2018-10-08& f090&14.8&7.8&5.9&24.6&4.4&4.9\\
        &&& f150&18.0&6.1&15.4&21.7&6.6&6.8\\
        &&& f220&66.7&42.8&51.3&80.4&37.5&34.3\\
        \hline
        \multirow{3}{*}{AT2019ppm}&88.0475,-7.4564&2019-09-07& f090&$43.9\pm8.5$&34.1&11.4&12.2&11.3&7.9\\
        &&& f150&$47.2\pm9.7$&36.1&21.2&34.8&23.1&23.0\\
        &&& f220&96.4&175.9&85.4&97.6&78.2&72.1\\
        \hline
        \multicolumn{10}{c}{\footnotesize This table is published in its entirety in the machine-readable format online. A portion is shown here for guidance regarding its form and content.}
    \end{tabular}
\end{table*}

\begin{table*}
\caption{Examples of the upper limits on flux density for TDEs. The columns correspond to the time range of the map in days, with 0 being the discovery time. All measurements are 95 per cent upper limits in mJy. Positions are in degrees. \label{table:tdes}}
\centering
    \begin{tabular}{  l  l  l  l  r  r  r  r  r  r  }
        \hline
        & & & & \multicolumn{6}{c}{Time interval for the map relative to alert (in days)} \\
        \colheadd{TDE name} & \colheadd{RA/Dec} & \colheadd{Discovery} & \colheadd{Freq.} & \colheadd{[$-$28,0]} & \colheadd{[0,28]} & \colheadd{[28,56]} & \colheadd{[56,84]} & \colheadd{[0,56]} & \colheadd{[0,84]} \\
         & deg. & & & mJy & mJy & mJy & mJy & mJy & mJy \\
        \hline
        \hline
        \multirow{3}{*}{AT2018fyk}&342.5670,-44.8649&2018-09-08& f090& ---&10.3&7.6&22.1&5.6&7.1\\
        &&& f150& ---&13.2&11.1&14.0&7.9&6.7\\
        &&& f220& ---&37.4&139.8&28.6&45.5&24.6\\
        \hline
        \multirow{3}{*}{AT2019qiz}&71.6578,-10.2264&2019-09-24& f090&17.5&14.1&19.6&31.1&14.1&15.4\\
        &&& f150&18.4&26.9&25.5&19.0&23.9&18.4\\
        &&& f220&116.2&64.8&92.3&62.0&69.1&46.2\\
        \hline
        \multirow{3}{*}{J013204.6+122236}&23.0187,12.3766&2020-07-08& f090&29.8&40.8&38.3&16.5&32.0&18.4\\
        &&& f150&43.2&26.2&15.9&23.1&13.2&12.2\\
        &&& f220&58.8&63.6&40.4&73.5&32.4&36.9\\
        \hline
        \multicolumn{10}{c}{\footnotesize This table is published in its entirety in the machine-readable format online. A portion is shown here for guidance regarding its form and content.}
    \end{tabular}
\end{table*}

\begin{table*}
\caption{Examples of the upper limits on flux density for GRBs. The columns correspond to the time range of the map in days, with 0 being the discovery time. The numbers are 95 per cent upper limits and in mJy. The position is in degrees. \label{table:grbs}}
\centering
\resizebox{\textwidth}{!}{    
    \begin{tabular}{  l  l  l  l  r  r  r  r  r  r  r  r  r  r  r  r  }
        \hline
        &  & & & \multicolumn{12}{c}{Time interval for the map relative to alert (in days)} \\
        \colheadd{GRB Name} & \colheadd{RA/Dec} & \colheadd{Discovery} & \colheadd{Freq.}  & \colheadd{[$-$3,0]} & \colheadd{[$-$7,0]} & \colheadd{[0,3]} & \colheadd{[3,6]} & \colheadd{[6,9]} & \colheadd{[9,12]} & \colheadd{[12,15]} & \colheadd{[0,7]} & \colheadd{[7,14]} & \colheadd{[14,21]} & \colheadd{[21,28]} & \colheadd{[0,28]} \\
         & deg. & & & mJy & mJy & mJy & mJy & mJy & mJy & mJy & mJy & mJy & mJy & mJy & mJy \\
        \hline
        \hline
        \multirow{1}{*}{131031A}&29.6102,-1.5788&2013-10-31& f150&10.1&13.2&24.1&14.6&19.9&11.0&6.8&15.1&8.5&5.3&5.0&4.1\\
        \hline
        \multirow{2}{*}{150710A}&194.4705,14.3181&2015-07-10& f090&72.0&43.7&82.4&---&36.0&25.5&72.0&51.7&20.6&34.5&17.6&18.5\\
        &&& f150&30.1&22.9&53.5&---&26.4&48.3&30.1&24.1&27.4&37.3&33.1&18.3\\
        \hline
        \multirow{3}{*}{171027A}&61.6907,-2.6221&2017-10-27& f090&---&20.4&31.4&22.3&---&23.1&---&13.1&25.1&12.9&12.9&6.0\\
        &&& f150&---&75.5&27.7&34.3&---&16.4&---&20.4&35.9&16.5&22.2&11.0\\
        &&& f220&---&322.6&146.8&127.2&---&144.2&---&133.3&156.8&63.6&92.8&64.9\\
        \multirow{3}{*}{171027A}&61.6907,-2.6221&2017-10-27& f090&---&20.4&31.4&22.3&---&23.1&---&13.1&25.1&12.9&12.9&6.0\\
        &&& f150&---&75.5&27.7&34.3&---&16.4&---&20.4&35.9&16.5&22.2&11.0\\
        &&& f220&---&322.6&146.8&127.2&---&144.2&---&133.3&156.8&63.6&92.8&64.9\\
        \hline
        \multirow{3}{*}{191004B}&49.2042,-39.6348&2019-10-04& f090&16.8&29.1&51.7&17.5&42.5&18.5&11.4&15.8&15.6&13.9&46.4&9.4\\
        &&& f150&28.5&56.2&35.2&27.2&61.2&23.1&21.2&20.9&19.6&40.1&23.1&15.6\\
        &&& f220&109.8&103.3&124.4&---&177.6&---&76.9&60.0&177.6&86.6&91.9&42.2\\
        \hline
        \multicolumn{16}{c}{\footnotesize This table is published in its entirety in the machine-readable format online. A portion is shown here for guidance regarding its form and content.}
    \end{tabular}
}
\end{table*}

\subsection{AT2019ppm} \label{sec:AT2019ppm}

While this paper does not target AGNs, one of the sources in our SNe/ATs catalogue seems to be a flare from AGN activity. AT2019ppm was reported on September 7, 2019 \citep{2019TNSTR1763....1N}. Its position coincides with the galaxy NGC~2110, which is also a known Seyfert II galaxy. The transient processing pipeline that reported the event, AMPEL \citep{2019A&A...631A.147N} at the Zwicky Transient Facility \citep{2019PASP..131g8001G}, flags a discovery when it is associated to a ``Known SDSS and/or MILLIQUAS QSO/AGN''. \footnote{\url{https://www.wis-tns.org/object/2019ppm}; see also \citet{2017A&A...597A..79P}} 

\begin{figure}
    \centering
    \includegraphics[width=\columnwidth]{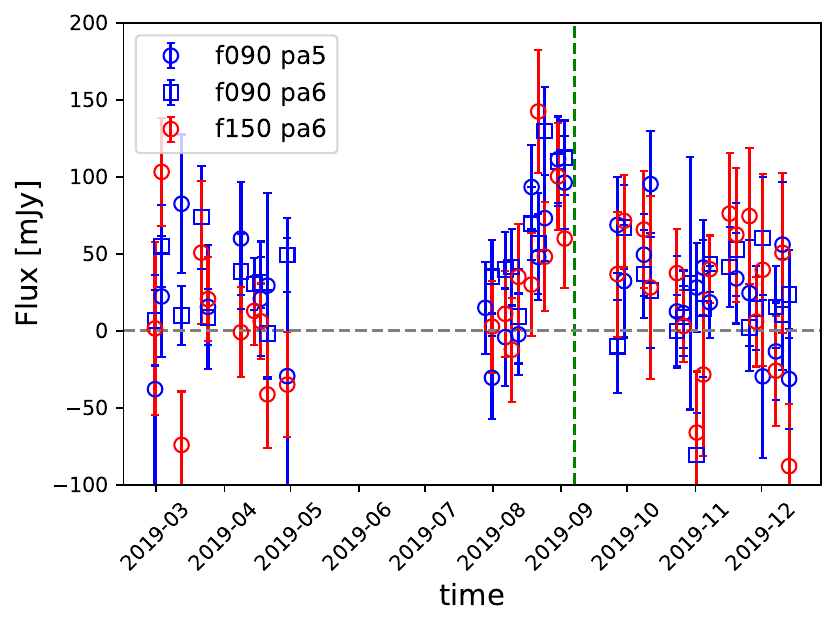}
    \caption{The light curve at the coordinates of the AT2019ppm event is shown using data from ACT’s 2019 observing season in the f090 PA5, f090 PA6, and f150 PA6 channels. We only plot one array for f150 so as not to overcrowd the plot. Each data point corresponds to three days of observations. The dashed green line shows the time when AT2019ppm was reported. These are total flux measurements (i.e., the flux of the background map has not been subtracted). The error bars are $1\sigma$ estimates from the matched filter (equation~\ref{eq:flux-dflux}).}
    \label{fig:AT2019ppm}
\end{figure}

All f090 and f150 arrays measure a consistently high signal in the centre of the map in the month before discovery. We measure an excess flux of $44\pm9$\,mJy for f090 and $47\pm10$\,mJy for f150 in this time interval, where we have averaged the fluxes from all arrays at a given frequency; note that this yields a S/N~$\approx 5$ for each frequency. The f220 flux does not seem to be significant ($41\pm33$\,mJy, which gives a 95 per cent upper limit of 96\,mJy). Light curves for f090 and f150  during ACT's 2019 season at the AT2019ppm coordinates are shown in Figure~\ref{fig:AT2019ppm}. The dashed green vertical line shows the discovery time of the transient. Each measurement corresponds to three days of data, and we have only included a single array at f150 so as not to overcrowd the plot. The light curves clearly show rising emission in the month previous to the discovery date. Unfortunately, we have a gap in our observations after the reported date of the AT2019ppm event, presumably during the period where the light curve would have peaked. ACT next observed these coordinates on September 25, 2019, at which point there is no longer any clear excess and when the transient's flux would presumably have returned to its regular level. Since the AT2019ppm transient is associated with NGC~2110, we conclude that we have detected a flare in the host galaxy's AGN that lasted for approximately one month.
Since the characterisation of the variability of AGNs in the ACT data is a topic of ongoing investigation in the collaboration, we do not further consider them in this work.

\subsection{Stacking maps across sources} \label{sec:stacking}

To increase the S/N of a potential detection, we stack multiple events together, in the same time intervals and at the same frequencies, for a common type of transient. We stack over luminosities, rather than fluxes, to obtain physically meaningful results. We do an inverse variance-weighted stack of luminosities, using the transformation $\vect{L}_i = 4 \pi D_i^2 \vect{f}_i$, where $D_i$ is the luminosity distance to source $i$. The stacked luminosity $\hat{\vect{L}}$ is then given by
\begin{equation}
    \hat{\vect{L}} = \frac{\sum_i \vect{\kappa}_{\rm L, i} \vect{L}_i}{\sum_i \vect{\kappa}_{\rm L, i}} \text{,}
\end{equation}
where $i$ iterates over the individual sources being stacked, and $\vect{\kappa}_{\rm L, i}$ is the inverse variance luminosity map for source $i$.
To increase the S/N further, we also stack sources across frequency channels, across time intervals, and across both. We combine time intervals and/or frequencies with the same simple inverse variance-weighted average of luminosities.\footnote{This might be improved by weighting each bin according to its expected signal strength based on a fiducial light curve and/or spectral energy distribution, and hence using a model. We avoid doing this here to keep the results easier to interpret.} Our results are reported in terms of characteristic luminosity $\nu L_{\nu}$, multiplying by the central frequency of each frequency channel.
In the following paragraphs, we briefly describe how we implement this stacking method for each of our three transient types.

We use the luminosity distance to the SN given by its measured redshift. The measured redshift distribution for all SNe and ATs is shown in Figure~\ref{fig:properties}. We separate our stack into two groups: one for core-collapse SNe and one for thermonuclear type Ia SNe. Radio/mm emission from core-collapse SNe probably depends on the progenitor and the interaction of the blast wave with the surrounding medium, and positive mm observations have been made of some types. On the other hand, thermonuclear type Ia SNe happen through a different process that is not expected to generate significant mm emission, and, to date, no mm detection has been made \citep[e.g.,][]{2016ApJ...821..119C,2020ApJ...890..159L}.\footnote{However, recently radio emission has been detected for the first time for a type Ia SN \citep{2023Natur.617..477K}.} Thus, we only consider spectroscopically-classified SNe of known types and perform the two separate stacks.

In the case of TDEs, we use the distance determined from measured redshifts (their distribution is shown in Figure~\ref{fig:properties}). We exclude 2 out of 12 sources which are possible, but unconfirmed, TDEs.

In the case of GRBs, we lack redshift measurements for most of our sample (measurements exist for only 24 out of 88 GRBs). We make a stack with only the GRBs that have a measured redshift. The redshifts vary in the range $0.14 \lesssim z \le 3.5$.

\subsubsection{Stacking Results}\label{sssec:stack_results}

The measured characteristic luminosity of the stacked maps are listed in Tables~\ref{table:stack_sne_CC}, \ref{table:stack_sne_Ia}, \ref{table:stack_tdes}, and \ref{table:stack_grbs} for core collapse SNe, type Ia SNe, TDEs, and GRBs, respectively.
In the tables, the rightmost column shows the maps stacked across time intervals, the bottom row shows the maps stacked across frequency channels, and the bottom right corner shows the overall stack for all time intervals and frequencies.

All of the measurements in the stacked maps show non-detections. For core-collapse SNe (Table~\ref{table:stack_sne_CC}), we constrain our stack to have [f090/f150/f220] characteristic luminosity less than [30/36/257]\,$\times 10^3$\,$L_{\sun}$ (at 95 per cent confidence level) when measured in the first 28 days of its discovery. For type Ia SNe (Table~\ref{table:stack_sne_Ia}), we constrain our stack to have [f090/f150/f220] characteristic luminosity less than [39/90/525]\,$\times 10^3$\,$L_{\sun}$ (at 95 per cent confidence level) when measured in the first 28 days of its discovery. For TDEs (Table~\ref{table:stack_tdes}), we constrain our stack to have [f090/f150/f220] characteristic luminosity less than [1.2/2.3/13.8]\,$\times 10^6$\,$L_{\sun}$ (at 95 per cent confidence level) when measured in the first 28 days of its discovery. For GRBs (Table~\ref{table:stack_grbs}), we constrain our stack to have [f090/f150/f220] characteristic luminosity less than [3.7/2.9/396]\,$\times 10^9$\,$L_{\sun}$ (at 95 per cent confidence level) when measured in the first three days of its discovery. We conclude that our stacked maps do not contain evidence of any source detection. As a reference, Figure~1 of \cite{2022ApJ...935...16E} shows examples of mm luminosity light curves for a wide range of extragalactic transients, including TDEs, FBOTs, core-collapse SNe, and LGRBs, in erg\,s$^{-1}$. The range of detections goes from $\sim 500$\,$L_{\sun}$ for the faintest SN examples to $\sim 3\times10^{10}$\,$L_{\sun}$ for the brightest examples of LGRBs. However, this sample of mm transients is certainly biased towards the brightest objects. While our upper limits are in theory below the luminosities of the brightest objects, in practice we are not stacking exceptionally bright transients. In other words, if every object we include in our stack was an exceptionally bright example, then we would have a positive detection in the stack at a very high significance, but that is not our case. This indicates that we do not reach the necessary depth to detect common examples of the different transient classes, at least with current ACT sensitivity, while we would be able to detect the very few and brightest examples if observed.

%%%%%%%%%%%%%%%%%%%%
% SNe core collapse
%%%%%%%%%%%%%%%%%%%%
\begin{table*}
    \caption{Characteristic luminosity for stacked core collapse SNe. The units are $10^3$\,$L_{\sun}$. The stacks use all arrays with the same frequency in a given time interval. The rightmost column shows the stacking across time intervals. The bottom row shows the stacking across frequency channels. The bottom right corner shows the stacking across both. \label{table:stack_sne_CC}}
    \begin{closetabcols}[1.5mm]
    \begin{center}
    \begin{tabular}{|c|rr|rr|rr|rr|rr|}
        \hline
        \hline
        \colheadd{Band} & \multicolumn{2}{c|}{\colheadd{days [$-$28,0]}} & \multicolumn{2}{c|}{\colheadd{days [0,28]}} & \multicolumn{2}{c|}{\colheadd{days [28,56]}} & \multicolumn{2}{c|}{\colheadd{days [56,84]}} & \multicolumn{2}{c|}{\colheadd{across time}} \\
                & $\nu L_{\nu}$  &  err  &   $\nu L_{\nu}$    &  err      & $\nu L_{\nu}$  &err     &$\nu L_{\nu}$   &err     &$\nu L_{\nu}$   &err    \\
        \hline
        \colheadd{f090}&3.6&15.2&6.3&12.8&-8.2&11.1&27.6&16.5&4.1&6.7\\ 
        \colheadd{f150}&-50.9&25.7&-9.5&21.0&-1.3&18.9&-21.8&27.0&-16.6&11.2\\ 
        \colheadd{f220}&349.0&145.1&-2.8&132.2&-29.6&103.8&29.2&146.8&61.6&64.0\\ 
        \hline
        \colheadd{Across freqs.}&-18.5&26.1&1.0&21.7&-12.1&19.2&19.3&27.8&-4.4&11.5\\ 
        \hline
    \end{tabular}
    \end{center}
    \end{closetabcols}
\end{table*}

%%%%%%%%%%%%%%%%%%%%
% SNe type Ia
%%%%%%%%%%%%%%%%%%%%
%\noindent \makebox[\textwidth]{ % This trick helps get the centering right.
\begin{table*}
    \caption{Characteristic luminosity for stacked type Ia SNe. The units are $10^3$\,$L_{\sun}$. 
    \label{table:stack_sne_Ia}}
    \hspace{-1.25in} % For some reason need to manually centre …
    \begin{closetabcols}[0.5em]
    \begin{center}
    \begin{tabular}{|c|rr|rr|rr|rr|rr|}
            \hline
            \hline
            \colheadd{Band} & \multicolumn{2}{c|}{\colheadd{days [$-$28,0]}} & \multicolumn{2}{c|}{\colheadd{days [0,28]}} & \multicolumn{2}{c|}{\colheadd{days [28,56]}} & \multicolumn{2}{c|}{\colheadd{days [56,84]}} & \multicolumn{2}{c|}{\colheadd{across time}} \\
                    & $\nu L_{\nu}$  &  err  &   $\nu L_{\nu}$    &  err       & $\nu L_{\nu}$  &err     &$\nu L_{\nu}$   &err      &$\nu L_{\nu}$   &err    \\
            \hline
            \colheadd{f090}&-2.9&20.9&-8.0&22.1&19.4&25.5&-2.1&29.3&0.6&11.9\\ 
            \colheadd{f150}&6.8&32.6&28.8&35.0&47.3&37.7&-55.1&41.2&10.1&18.1\\ 
            \colheadd{f220}&331.6&167.8&107.2&228.7&264.5&189.5&13.7&266.8&221.6&101.8\\
            \hline
            \colheadd{Across freqs.}&15.5&34.4&14.4&36.9&69.2&40.8&-46.5&45.9&16.3&19.4\\ 
            \hline
    \end{tabular}
    \end{center}
    \end{closetabcols}
\end{table*}
%}

%%%%%%%%%%%%%%%%%
% TDEs
%%%%%%%%%%%%%%%%%

\begin{table*}
    \caption{Characteristic luminosity for stacked TDEs. The units are $10^6$\,$L_{\sun}$. 
    \label{table:stack_tdes}}
    \begin{closetabcols}[1.5mm]
    \begin{center}
    \begin{tabular}{|c|rr|rr|rr|rr|rr|}
        \hline
        \hline
        \colheadd{Band} & \multicolumn{2}{c|}{\colheadd{days [$-$28,0]}} & \multicolumn{2}{c|}{\colheadd{days [0,28]}} & \multicolumn{2}{c|}{\colheadd{days [28,56]}} & \multicolumn{2}{c|}{\colheadd{days [56,84]}} & \multicolumn{2}{c|}{\colheadd{across time}} \\
                & $\nu L_{\nu}$  &  err  &   $\nu L_{\nu}$    &  err       & $\nu L_{\nu}$  &err      &$\nu L_{\nu}$   &err       &$\nu L_{\nu}$   &err    \\
        \hline
        \colheadd{f090}&-1.48&1.12&-1.02&0.89&0.99&0.66&2.29&0.88&0.49&0.42\\ 
        \colheadd{f150}&-2.87&1.77&0.08&1.14&1.52&1.09&-0.44&1.54&0.11&0.65\\ 
        \colheadd{f220}&15.85&9.71&-3.06&7.99&14.17&6.66&-3.39&7.80&5.88&3.91\\ 
        \hline
        \colheadd{Across freqs.}&-3.21&1.87&-0.97&1.33&2.67&1.12&2.52&1.53&0.83&0.69\\ 
        \hline
        \end{tabular}
    \end{center}
    \end{closetabcols}
\end{table*}

%%%%%%%%%%%%%%%%%
% GRBs
%%%%%%%%%%%%%%%%%

\begin{table*}
    \caption{Characteristic luminosity for stacked GRBs. The units are $10^9$\,$L_{\sun}$.
    \label{table:stack_grbs}}
    \scriptsize
    \begin{closetabcols}[0.5em]
    \begin{center}
    \begin{tabular}{|c|rr|rr|rr|rr|rr|rr|rr|rr|rr|rr|}
        \hline
        \hline
        \colheadd{Band} & \multicolumn{2}{c|}{\colheadd{days [$-$7,0}]} & \multicolumn{2}{c|}{\colheadd{days [$-$3,0]}} & \multicolumn{2}{c|}{\colheadd{days [0,3]}} & \multicolumn{2}{c|}{\colheadd{days [3,6]}} & \multicolumn{2}{c|}{\colheadd{days [6,9]}} & \multicolumn{2}{|c|}{\colheadd{days [9,12]}} & \multicolumn{2}{c|}{\colheadd{days [12,15]}} & \multicolumn{2}{c|}{\colheadd{days [14,21]}} & \multicolumn{2}{c|}{\colheadd{days [21,28]}} & \multicolumn{2}{c|}{\colheadd{across time}} \\
                & $\nu L_{\nu}$  &  err  & $\nu L_{\nu}$  &  err & $\nu L_{\nu}$  &  err  & $\nu L_{\nu}$  &  err  & $\nu L_{\nu}$  &  err & $\nu L_{\nu}$  &  err  & $\nu L_{\nu}$  &  err  & $\nu L_{\nu}$  &  err  & $\nu L_{\nu}$  &  err  & $\nu L_{\nu}$  &  err  \\
        \hline
        \colheadd{f090}&-0.12&0.34&-3.23&1.35&1.13&1.38&3.82&11.69&2.04&0.98&0.59&1.42&2.46&4.76&1.18&1.26&0.42&3.59&0.09&0.29\\ 
        \colheadd{f150}&-0.17&0.44&-3.10&2.71&-6.17&2.77&-39.14&17.56&-0.82&2.05&-3.04&2.71&-27.19&27.57&1.41&2.36&-0.20&6.32&-0.44&0.40\\ 
        \colheadd{f220}&-15.01&102.43&-35.06&105.94&-160.93&250.04&67.08&83.62&190.46&167.84&7.04&83.60&-63.28&113.91&-42.27&121.91&37.66&89.34&9.13&35.54\\ 
        \hline
        \colheadd{Across freqs.}&-0.28&0.51&-6.58&2.53&-1.84&2.58&-21.71&18.79&2.69&1.87&-1.00&2.61&2.07&10.85&2.52&2.30&0.62&6.37&-0.27&0.46\\                 
        \hline        
    \end{tabular}   
    \end{center}
    \end{closetabcols}
\end{table*}
\section{Discussion and conclusions} \label{sec:conclusions}

We have presented a search for excess flux in ACT data before and after the discovery of 203 SNe and ATs, 12 TDEs, and 88 GRBs. We make no significant detection of excess flux (S/N~$< 4$ for almost all cases; see Figure~\ref{fig:histogram}), except for a single source, AT2019ppm, that seems to be explained by AGN activity of the host galaxy (Section~\ref{sec:AT2019ppm}). We place upper limits for all time intervals considered before and after the transients included in our search at the f090, f150, and f220 ACT frequency channels; the full results are available as supplementary material. We increased the S/N by stacking maps of multiple sources of the same class, with stacking weights based on the luminosity of the expected source, but still do not achieve a significant enough S/N for any detection. 

In conclusion, our non-detection of any extragalactic transient source is not unexpected, given that ACT lacked the sensitivity required to detect most of the sources. Not all the sources we observe are expected to produce mm emission, and those that are should typically only produce mm emission at the sub-mJy level, which is not possible for ACT to measure on short time scales. Rare, very energetic events such as AT2018cow could have been detected at high significance by ACT (AT2018cow itself was located just north of our survey's maximum declination and also occurred at a time when the telescope was idle). While our upper limits are too high to provide insight into the physics of the transients considered, this work represents an early effort in the new field of time domain science using mm survey data and introduces methodologies that should be useful for future transient searches. Archival searches like the one in this paper are crucially limited by the sensitivity of individual observations. This can be overcome by stacking sources, but several hundreds of sources might ultimately be required, and stacking comes with the multiple caveats that have been mentioned in this paper, especially if sources are dissimilar. Future experiments, such as SO and CMB-S4, will have the capability of potentially detecting hundreds of on-axis LGRBs in addition to tens of other events such as FBOTs and on-axis TDEs \citep{2022ApJ...935...16E}. For example, SO's large aperture telescope (LAT) will have about 4.7 times the number of detectors as ACT did in the f090, f150 and f220 bands when PA4–6 were all installed,\footnote{Specifically, SO will have 10,320 detectors at 93\,GHz, 10,320 at 145\,GHz and 5,160 at 225\,GHz \citep{2021ApJS..256...23Z}, compared to 1,712 at f090, 2,718 at f150 and 1,006 at f220 when all of PA4--6 were installed \citep{2016JLTP..184..772H}. In the field, not all detectors work (see, e.g., \citealt{2020JCAP...12..047A}), but here we assume that the SO yields will be similar to ACT's.} corresponding to a ${\sim}2.2\times$ increase in sensitivity to transient events if all frequencies are equally weighted. Furthermore, over its five year campaign, we expect SO to spend roughly a factor of three times longer covering the same ${\sim}40$ per cent of the sky than was analysed in this paper, for a further ${\sim}1.75\times$ increase in sensitivity.\footnote{ACT observed ${\sim}40$ per cent of the sky from 2016--2021 (we did not include 2022 in this paper), which is five years, but we had significant stoppages due to telescope failures and upgrades, and also did not have a full complement of f090/150/220 on the sky every year (e.g., after 2019, PA6 was replaced with the PA7 f030/040 array). We thus adopt a rough estimate of ${\sim}50$ per cent better observing efficiency with SO. Furthermore, we assume that we will include daytime data in addition to nighttime data for a further ${\sim}2\times$ improvement, giving a total improvement of ${\sim}3\times$.} Thus, the stacking of sources as done in the same fashion as in this work would improve the detection threshold by a factor $\sim 3.85$. 
The recently-announced Advanced SO project will eventually double the number of detectors in the LAT and run an additional five years, which improves the uncertainty by a factor $\sim 6.7$ as compared to this work over the full lifetime of SO. Additional improvements in sensitivity would come with the inclusion of the SO LAT's 27, 39 and 280\,GHz channels \citep{2021ApJS..256...23Z}.

While the targets in our study were known, archival transients, the ACT collaboration is preparing several other time domain projects. We performed a systematic, optimised blind transient search in which we found 14 new candidates similar to those discovered serendipitously by \cite{2021ApJ...915...14N} \citep{2023ApJ...956...36L}, made a targeted search of asteroids in the ACT survey, detecting $\sim 160$ of them \citep{2023arXiv230605468O}, and another study characterising the variability of AGNs is underway. As CMB telescopes continue to deliver impressive cosmological results in the coming years, we can also look forward to a new era of time domain science thanks to their wide-area, high-sensitivity, and short-cadence observations.

\section*{Acknowledgements}

We thank the anonymous referee for very useful suggestions that improved this paper.
CHC and KMH acknowledge NSF award 1815887. KMH acknowledges NSF award 2206344. CHC acknowledges ANID FONDECYT Postdoc Fellowship 3220255. ADH acknowledges support from the Sutton Family Chair in Science, Christianity and Cultures, from the Faculty of Arts and Science, University of Toronto, and from the Natural Sciences and Engineering Research Council of Canada (NSERC) [RGPIN-2023-05014, DGECR-2023-00180]. This work was supported by a grant from the Simons Foundation (CCA 918271, PBL). CHC, RD, RP and CV thank BASAL CATA  FB210003. EC acknowledges support from the European Research Council (ERC) under the European Union’s Horizon 2020 research and innovation programme (Grant agreement No. 849169).

This  work  was  supported  by  the  U.S.  National  Science Foundation  through  awards  AST-0408698,  AST-0965625,  and  AST-1440226  for  the  ACT  project,  as well as awards PHY-0355328,  PHY-0855887 and PHY-1214379. Funding was also provided by Princeton University, the  University  of  Pennsylvania, and  a  Canada Foundation for Innovation (CFI) award to UBC.

ACT operates in the Parque Astron\'{o}mico Atacama in northern Chile under the auspices of the Agencia Nacional de Investigaci\'{o}n y Desarrollo (ANID; formerly Comisi\'{o}n Nacional de Investigaci\'{o}n Cient\'{i}fica y Tecnol\'{o}gica de Chile, or CONICYT). We thank the Republic of Chile for hosting ACT in the northern Atacama, and the local indigenous Licanantay communities whom we follow in observing and learning from the night sky.

The development of multichroic detectors and lenses was supported by NASA grants NNX13AE56G and NNX14AB58G. Detector research at NIST was supported by the NIST Innovations in Measurement Science program. Computations were performed on Cori at NERSC as part of the CMB Community allocation, on the Niagara supercomputer at the SciNet HPC Consortium, and on Feynman and Tiger at Princeton Research Computing, and on the hippo cluster at the University of KwaZulu-Natal. SciNet is funded by the CFI under the auspices of Compute Canada, the Government of Ontario, the Ontario Research Fund--Research Excellence, and the University of Toronto.

Colleagues at AstroNorte and RadioSky provide logistical support and keep operations in Chile running smoothly. We also thank the Mishrahi Fund and the Wilkinson Fund for their generous support of the project.

This research has made use of the NASA/IPAC Extragalactic Database (NED), which is funded by the National Aeronautics and Space Administration and operated by the California Institute of Technology. 

This research used data from Sloan Digital Sky Survey IV. Funding for the Sloan Digital Sky Survey IV has been provided by the Alfred P. Sloan Foundation, the U.S. Department of Energy Ofﬁce of Science, and the Participating Institutions. SDSS-IV acknowledges support and resources from the Center for High-Performance Computing at the University of Utah. The SDSS web site is http://www.sdss.org. SDSS-IV is managed by the Astrophysical Research Consortium for the Participating Institutions of the SDSS Collaboration, including the Brazilian Participation Group, the Carnegie Institution for Science, Carnegie Mellon University, the Chilean Participation Group, the French Participation Group, Harvard-Smithsonian Center for Astrophysics, Instituto de Astrofísica de Canarias, Johns Hopkins University, Kavli Institute for the Physics and Mathematics of the Universe (IPMU)/University of Tokyo, Lawrence Berkeley National Laboratory, Leibniz Institut für Astrophysik Potsdam (AIP), Max-Planck-Institut für Astronomie (MPIA Heidelberg), Max-Planck-Institut für Astrophysik (MPA Garching), Max-Planck-Institut für Extraterrestrische Physik (MPE), National Astronomical Observatory of China, New Mexico State University, New York University, University of Notre Dame, Observatário Nacional/MCTI, The Ohio State University, Pennsylvania State University, Shanghai Astronomical Observatory, United Kingdom Participation Group, Universidad Nacional Autónoma de México, University of Arizona, University of Colorado Boulder, University of Oxford, University of Portsmouth, University of Utah, University of Virginia, University of Washington, University of Wisconsin, Vanderbilt University, and Yale University.
Some of the results and plots in this paper have been produced with the following software: \textsc{matplotlib} \citep{software:matplotlib}; \textsc{NumPy} \citep{software:numpy}; \textsc{pixell};\footnote{\url{https://pixell.readthedocs.io/en/latest/index.html}} \textsc{SciPy} \citep{software:scipy}.

%%%%%%%%%%%%%%%%%%%%%%%%%%%%%%%%%%%%%%%%%%%%%%%%%%
\section*{Data Availability}
The upper limit flux measurements for SNe/ATs, TDEs, and GRBs will be available as machine readable tables on the NASA Legacy Archive Microwave Background Data Analysis (LAMBDA) website.~\footnote{\url{https://lambda.gsfc.nasa.gov/product/act/actadv_targeted_transient_constraints_2023_info.html}}

%The inclusion of a Data Availability Statement is a requirement for articles published in MNRAS. Data Availability Statements provide a standardised format for readers to understand the availability of data underlying the research results described in the article. The statement may refer to original data generated in the course of the study or to third-party data analysed in the article. The statement should describe and provide means of access, where possible, by linking to the data or providing the required accession numbers for the relevant databases or DOIs.


\bibliographystyle{mnras}
\bibliography{biblio} % if your bibtex file is called example.bib

%\begin{thebibliography}{99}
%\end{thebibliography}

%%%%%%%%%%%%%%%%%%%%%%%%%%%%%%%%%%%%%%%%%%%%%%%%%%

%%%%%%%%%%%%%%%%% APPENDICES %%%%%%%%%%%%%%%%%%%%%

\appendix
\section{Estimation of the calibration variance} \label{sec:calibration}

A key issue in time domain analyses with ACT or any future experiment is characterising how much the calibration changes with time. For ACT, temporal changes in calibration could be caused by variations in the telescope optics, thermal fluctuations in the detector arrays and changes in atmospheric loading and emission, to name a few. The calibration stability can be estimated using an object whose time dependent flux is sufficiently well known a priori, since any deviations from the known flux can be attributed to calibration errors.

We use observations of Uranus to estimate our calibration variance. Uranus was frequently observed throughout the ACTPol and AdvACT data-taking seasons, and is bright enough to provide high SNR in a single observation but not so bright so as to create a non-linear detector response. In addition to Uranus, we considered using the light curves of objects expected to have a constant intrinsic flux, including the brightest dusty point sources identified in the f220 coadded ACT map \citep{2020JCAP...12..046N}, the core of the Orion nebula, the core of the M87 galaxy, and the extended radio lobes of galaxies with AGNs and jets. Each of these sources turned out to be more difficult to use than Uranus, either because they have small SNR or are extended and tricky to model as point-like sources. We therefore defer characterising the ACT calibration with these and other sources to a future study.

ACT made targeted observations of Uranus every few days as part of its routine calibration program, which uses a model of the planet's brightness to convert power incident on the detectors to brightness temperature units (pW to \si{\micro \kelvin}; \citealt{2020JCAP...12..047A}).\footnote{The final absolute calibration of the ACT maps uses the \Planck maps which are calibrated based on the dipole induced by Earth's solar period. Uranus is used as an intermediate, intraseason calibrator. The variation in the calibration factor ranges between 1 and 6 per cent depending on array and frequency.}  In addition to these targeted observations we used any chance observations of Uranus when it happened to be in the field during regular CMB observations.

\begin{figure*}
    \centering
    \includegraphics[width=1.0\textwidth]{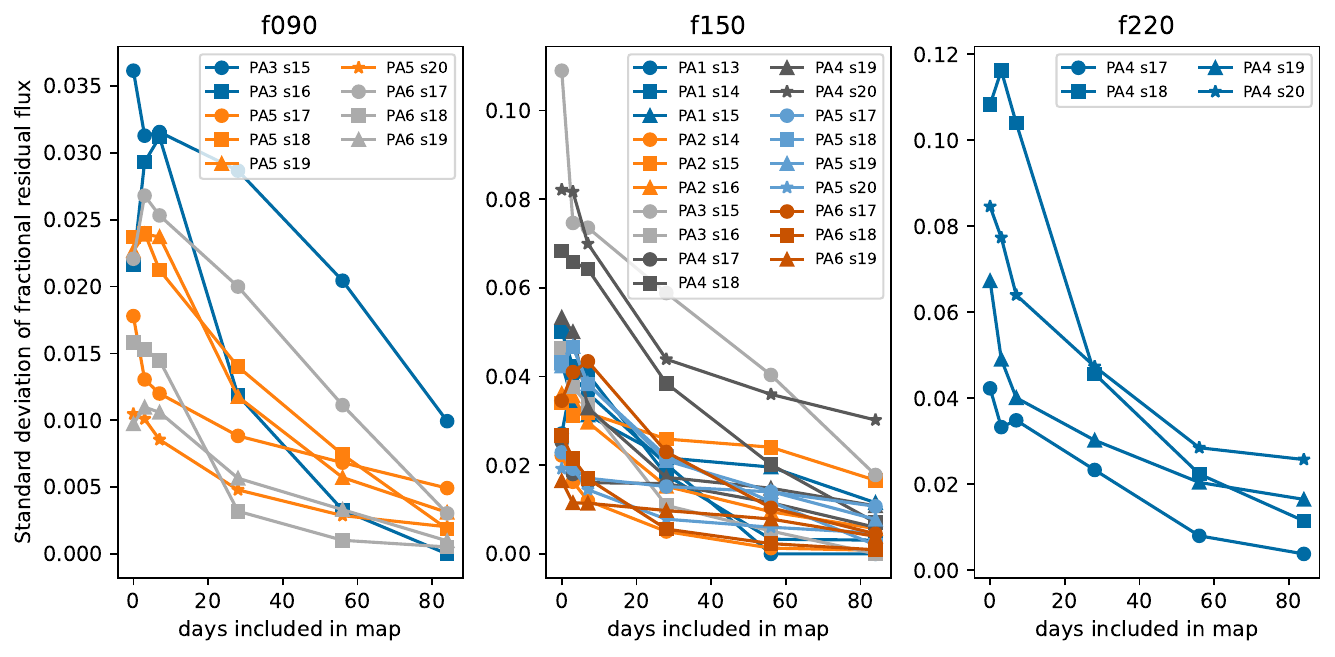}
    \caption{Standard deviation of the fractional residual flux of Uranus for every frequency and array. We bin the Uranus light curve (which has individual observation) into 3-, 7-, 28-, 56-, or 84-day bins and calculate the standard deviation. The standard deviation at zero days included in the map corresponds to the individual observations light curve, before binning.}
    \label{fig:calibration}
\end{figure*}

We construct light curves of Uranus for every season, frequency channel and detector array, obtaining fluxes using the matched filter described in Section~\ref{sec:matched-filter} and normalising them by the distance between Uranus and Earth. For each of these light curves we subtract its mean flux, and then divide by this mean to obtain a light curve in units of fractional residual flux. Finally, we bin it in in time intervals corresponding to the different maps we use for our sources; as detailed in Section~\ref{sec:selection}: 3, 7, 28, 56, and 84 days. We create 4,000 random intervals along the length of a season, and therefore a given observation will fall inside many of these intervals. We calculate the standard deviation of the fractional residual flux over these 4,000 intervals at every frequency channel, detector array, and season. These are shown in Figure~\ref{fig:calibration}. As the light curve is binned into longer time intervals, the standard deviation decreases. The f090 channel has the most stable calibration, with standard deviations around a couple of percent for the three-day maps, while the f220 channel is the noisiest, reaching more than 10 per cent for the three-day maps. We also include the standard deviation of the non-binned, full light curves: this is shown at zero days included in map.

\begin{figure*}
    \centering
    \includegraphics[width=0.49\textwidth]{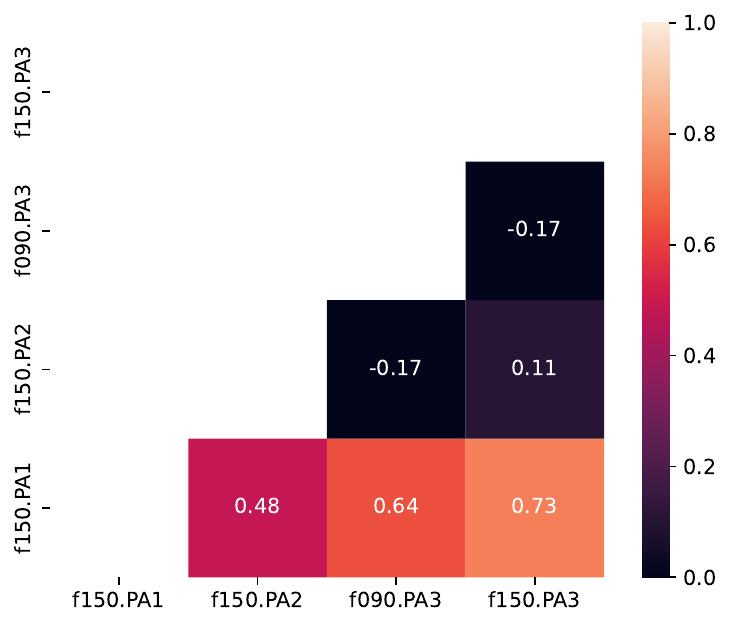}
    \includegraphics[width=0.49\textwidth]{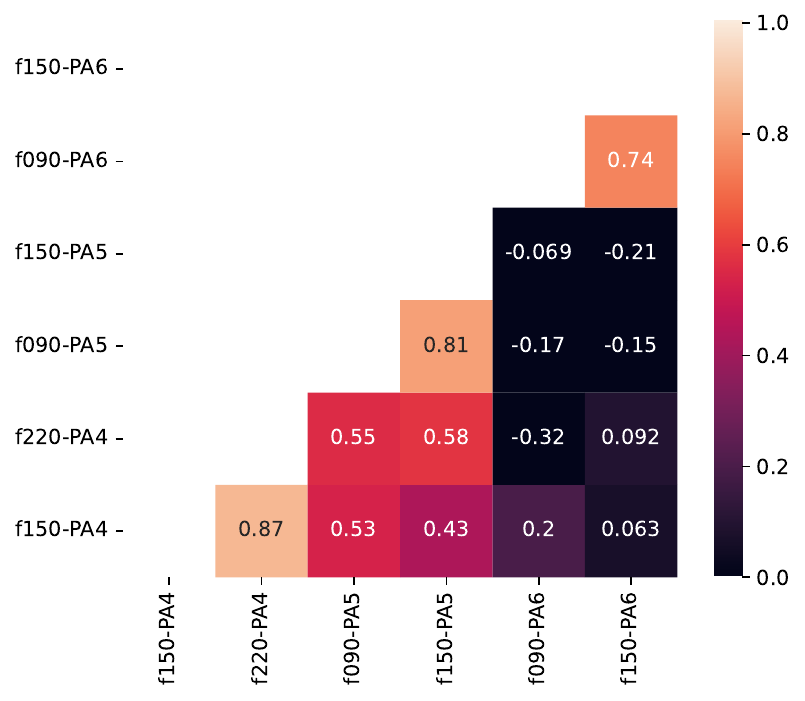}
    \caption{Pearson correlation coefficient matrix for the Uranus fractional residual light curves. The light curve from a frequency and array is correlated in time (observations taking place simultaneously) with another. On the left, the correlation for ACTPol (PA1, 2, 3), on the right the correlation for AdvACT (PA4, 5, 6).}
    \label{fig:correlation}
\end{figure*}

The nature and cause of the gain fluctuations measured above can be probed by measuring the correlation between fractional residual light curves from different arrays and/or frequencies. Since the telescope scans in azimuth at a constant elevation, the two bottom arrays in the focal plane (PA1 and PA2 for ACTPol, and PA4 and PA5 for AdvACT) observe the same sky position almost simultaneously (within a few seconds of each other), whereas the top (PA3 in ACTPol and PA6 in AdvACT) will observe the same sky position at a minimum of 4 minutes earlier (later), after the sky has rotated into (out of) its field of view. Gain variations due to drifts in the detector temperatures or changes in loading from the atmosphere might therefore be manifested in differences in the correlation between the pair of arrays in the bottom row and pairs of arrays split between the top and bottom row. To test this, we bin the Uranus light curves for each array/frequency channel into one day intervals and calculate the Pearson correlation coefficient for every possible pair of light curves, only retaining bins in which both light curves in the pair had measurements.

Figure~\ref{fig:correlation} shows the Pearson correlation coefficient for all combinations of frequencies and arrays, grouped into the ACTPol arrays (PA1, 2, 3; left panel) and the AdvACT arrays (PA4, 5, 6; right panel). For AdvACT, we observe a relatively high correlation between the two frequency channels in the same array. Correlations between PA4 and PA5 are also relatively high, reflecting the fact that these two arrays observe the same coordinates nearly simultaneously. However, the correlations between either PA4 or PA5 and PA6 are typically much lower. This is evidence that significant gain variations can occur on sub-day time scales, and that some of the gain variance shown in Figure~\ref{fig:calibration} is from short term fluctuations rather than longer term drifts. For ACTPol (left panel of Figure~\ref{fig:correlation}), this pattern is less obvious (e.g., both frequencies of PA3 exhibit substantial correlation with PA1, but not with PA2). However, there are many fewer data available for ACTPol and we might not have sufficient statistics to draw conclusions. We have found evidence consistent with these results using AGN light curves rather than Uranus; this work (in preparation) will have more robust statistics since it uses scores of sources.

The above correlation test suggests that flux variations can be attributed to systematics, but we should also consider the possibility that there is intrinsic flux variation in Uranus itself. \cite{2013ApJS..209...17H} analysed ACT observations of Uranus and compared them with the model of \cite{1993Icar..105..537G}, finding general agreement. The brightness of  Uranus is probably latitude dependent, but since its sub-Earth latitude  changes slowly, the mean brightness across the observed disk will vary on time scales of decades \citep[e.g.,][]{2008A&A...482..359K} and we expect that it will remain approximately constant for the duration of an ACT season. Any flux variation is therefore expected to be dominated by the changing Earth-Uranus distance, which we know precisely and account for.

Effectively, we now have a new source of systematic error, a fractional gain error $g$ estimated as described above and shown in Figure~\ref{fig:calibration}. The combined new uncertainty in the flux measured for a transient map is the sum in quadrature $\Delta f'^2 = \Delta f^2 + (s g)^2$, where $\Delta f$ is the uncertainty measured directly in the map and $s$ is the expected signal. By the nature of the sources we are trying to measure, we might expect for the signal to be very small and therefore $\Delta f' \sim \Delta f$. However, we decide to use the more conservative approach where the expected signal is of the order of the $\Delta f$ uncertainty, leading to $\Delta f' = \Delta f \sqrt{1+g^2}$. Based on this analysis, we multiply each error bar calculated from our matched filter maps by the factor $\sqrt{1+g^2}$ for all the results reported in this work. 

\section{Effect of stripy noise}\label{sec:stripy-noise}

\begin{figure}
    \centering
    \includegraphics[width=1\columnwidth]{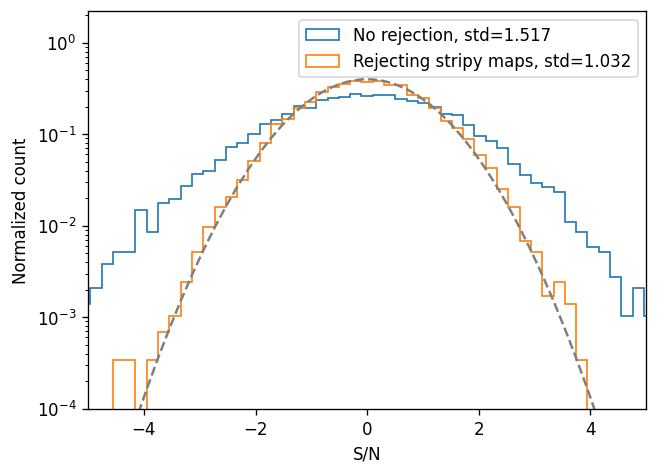}
    \caption{Histogram of pixels from the flux S/N map for the GRB stack at frequency f150 and time interval 12--15 days after discovery. The maps are a 1 degree square stamp with pixels inside a 2\,arcmin radius from the centre masked. The blue line is without rejecting any stripy maps, the orange line is after rejecting stripy maps. A normal distribution with standard deviation of 1 is shown as reference in dashed grey.
    }
    \label{fig:histogram_pixels}
\end{figure}

In this section, we briefly discuss the statistics of the stacked maps. In Section~\ref{sec:noisy-maps}, we described how some individual maps have anisotropic noise, visible as striping in the maps. Because our matched filter (Section~\ref{sec:matched-filter}) uses a noise model that assumes isotropy, the S/N of these maps would be significantly mis-estimated. To illustrate this with a pertinent example, consider the GRB stack for frequency channel f150 and time range between 12--15 days after discovery (c.f., Table~\ref{table:stack_grbs}). If we do not remove the stripy maps from our sample, 30 individual maps go into the stack and yield a characteristic luminosity of $(-27.50 \pm 9.44) \times 10^9$\,$L_{\odot}$, an apparent $-2.9\sigma$ measurement. However, Figure~\ref{fig:histogram_pixels} shows a histogram of the pixel values of the S/N map of the stack (wider, blue histogram) and demonstrates that the S/N has been significantly over-estimated. When 11 stripy maps are removed as per the method of Section~\ref{sec:noisy-maps}, the histogram of the S/N in the resulting stacked map (orange histogram in Figure~\ref{fig:histogram_pixels}) has a width consistent with the expected normal distribution. In this case, we measure characteristic luminosity of $(-27.19 \pm 27.57) \times 10^9$\,$L_{\odot}$, which corresponds to only ${\sim}-1\sigma$.

\section*{Affiliations}
\noindent\textit{
$^{1}$Department of Physics, Florida State University, Tallahassee, Florida 32306, USA\\
$^{2}$Instituto de Astrof\'isica and Centro de Astro-Ingenier\'ia, Facultad de F\'isica, Pontificia Universidad Cat\'olica de Chile, Av. Vicu\~na Mackenna 4860, 7820436 Macul, Santiago, Chile \\
$^{3}$Institute of Theoretical Astrophysics, University of Oslo, Norway \\
$^{4}$David A. Dunlap Department of Astronomy \& Astrophysics, University of Toronto, 50 St. George St., Toronto ON M5S 3H4, Canada \\
$^{5}$Specola Vaticana (Vatican Observatory), V-00120 Vatican City State \\
$^{6}$School of Physics and Astronomy, Cardiff University, The Parade, Cardiff, Wales CF24 3AA, UK \\
$^{7}$Department of Physics and Astronomy, University of Pennsylvania, 209 South 33rd Street, Philadelphia, Pennsylvania 19104, USA \\
$^{8}$Joseph Henry Laboratories of Physics, Jadwin Hall, Princeton University, Princeton, NJ 08544 \\
$^{9}$Department of Astrophysical Sciences, Princeton University, Princeton, New Jersey 08544, USA \\
$^{10}$Kavli Institute for Cosmological Physics, University of Chicago, 5640 S. Ellis Ave., Chicago, IL 60637, USA \\
$^{11}$Wits Centre for Astrophysics, School of Physics, University of the Witwatersrand, Private Bag 3, 2050, Johannesburg, South Africa \\
$^{12}$School of Mathematics, Statistics \& Computer Science, University of KwaZulu-Natal, Westville Campus, Durban 4041, South Africa \\
$^{13}$Department of Astronomy, Cornell University, Ithaca, NY 14853, USA \\
$^{14}$Department of Physics \& Astronomy, University of Victoria, Victoria, BC, V8P 1A1, Canada \\
$^{15}$Department of Physics, Cornell University, Ithaca, NY 14853, USA \\
$^{16}$Department of Physics and Astronomy, Haverford College, Haverford, PA, USA 19041 \\
$^{17}$Department of Physics, Stanford University, Stanford, California 94305, USA \\
$^{18}$Kavli Institute for Particle Astrophysics and Cosmology, Stanford, CA 94305, USA \\
$^{19}$Instituto de F\'isica, Pontificia Universidad Cat\'olica de Valpara\'iso, Casilla 4059, Valpara\'iso, Chile \\
$^{20}$Department of Physics, Jadwin Hall, Princeton University, Princeton, NJ 08544, USA \\
$^{21}$NASA Goddard Space Flight Center, 8800 Greenbelt Rd, Greenbelt, MD 20771, USA \\
}
%%%%%%%%%%%%%%%%%%%%%%%%%%%%%%%%%%%%%%%%%%%%%%%%%%

% Don't change these lines
\bsp	% typesetting comment
\label{lastpage}
\end{document}